\documentclass[aps,preprint]{revtex4-2}
\usepackage{amsmath}
\usepackage[english]{babel}
\usepackage{amssymb}
\usepackage{graphicx}
\usepackage{epsfig}
\usepackage{bm}
\usepackage{dcolumn}
\usepackage{color}

\usepackage{bbm}
\usepackage{soul}
\usepackage{verbatim}
\usepackage{ulem}

\def\gcoups{g_s}
\def\gcoupr{g_{v_3}}
\def\gcoupw{g_{v_0}}

\def\alphagen{a}
\def\betagen{b}

\def\gammagen{d}

\def\gcoupt{g}

\def\cc{\alpha}

\def\gmatrix{{\cal G}}

\def\jmatrix{{\cal J}}

\def\gnomatrix{{G}}

\def\jnomatrix{{J}}

\def\taua{\tau_a}
\def\sigmaa{\sigma_a}
\def\pia{\pi_a}
\def\rhoamu{\rho_{a\mu}}

\def\sigmac{\sigma_0}
\def\pic{\pi_0}
\def\rhocmu{\rho_{0\mu}}
\def\rhoc{\rho_{0}}

\def\sigmat{\sigma_3}
\def\pit{\pi_3}
\def\rhotmu{\rho_{3\mu}}
\def\rhot{\rho_3}

\def\pib{\pi_{b}}
\def\rhobmu{\rho_{b\mu}}
\def\rhobnu{\rho_{b\nu}}

\def\rhoapar{\rho_{a\parallel}}
\def\rhoaperp{\rho_{a\perp}}
\def\rhocpar{\rho_{0\parallel}}
\def\rhocperp{\rho_{0\perp}}
\def\rhotpar{\rho_{3\parallel}}
\def\rhotperp{\rho_{3\perp}}
\def\pif{\pi_f}
\def\rhof{\rho_f}

\def\piu{\pi_{u}}
\def\rhoupar{\rho_{u\parallel}}
\def\rhouperp{\rho_{u\perp}}

\def\pid{\pi_{d}}
\def\rhodpar{\rho_{d\parallel}}
\def\rhodperp{\rho_{d\perp}}

\def\trmin{{\rm tr}}
\def\mf{{\mbox{\tiny MF}}}

\DeclareMathOperator\arctanh{arctanh}
\DeclareMathOperator\arccoth{arccoth}
\DeclareMathOperator\arcsinh{arcsinh}

\begin{document}

\title{\sc{\Large Neutral pseudoscalar and vector meson masses
under strong magnetic fields in an extended NJL model: mixing effects}}

\author{J.P. Carlomagno$^{a,b}$, D. G\'omez Dumm$^{a,b}$, S. Noguera$^{c}$ and N.N.\ Scoccola$^{b,c,d}$}

\affiliation{$^{a}$ IFLP, CONICET $-$ Departamento de F\'{\i}sica, Facultad de Ciencias Exactas,
Universidad Nacional de La Plata, C.C. 67, (1900) La Plata, Argentina}
\affiliation{$^{b}$ CONICET, Rivadavia 1917, (1033) Buenos Aires, Argentina}
\affiliation{$^{c}$ Departamento de F\'{\i}sica Te\'{o}rica and IFIC, Centro Mixto
Universidad de Valencia-CSIC, E-46100 Burjassot (Valencia), Spain}
\affiliation{$^{d}$ Physics Department, Comisi\'{o}n Nacional de Energ\'{\i}a At\'{o}mica, \\
Av.\ Libertador 8250, (1429) Buenos Aires, Argentina}

\begin{abstract}
Mixing effects on the mass spectrum of light neutral pseudoscalar and vector
mesons in the presence of an external uniform magnetic field $\vec B$ are
studied in the framework of a two-flavor NJL-like model. The model includes
isoscalar and isovector couplings both in the scalar-pseudoscalar and vector
sectors, and also incorporates flavor mixing through a 't Hooft-like
term. Numerical results for the $B$ dependence of meson masses are compared
with present lattice QCD results. In particular, it is shown that the mixing
between pseudoscalar and vector meson states leads to a significant
reduction of the mass of the lightest state. The role of chiral symmetry and
the effect of the alignment of quark magnetic moments in the presence of the
magnetic field are discussed.
\end{abstract}
\date{\today}


\maketitle

\renewcommand{\thefootnote}{\arabic{footnote}}
\setcounter{footnote}{0}

\section{Introduction}

It is well known that the presence of a large background magnetic field
$\vec B$ has a significant impact on the physics of strongly interacting
particles, leading to important effects on both hadron properties and QCD
phase transition
features~\cite{Kharzeev:2012ph,Andersen:2014xxa,Miransky:2015ava}. By a
``large'' field it is understood here that the order of magnitude of $B$
competes with the QCD confining scale $\Lambda_{\rm QCD}$ squared, i.e.\
$|eB|\gtrsim \Lambda_{\rm QCD}^2$, $|B|\gtrsim 10^{19}$~G. Such huge
magnetic fields can be achieved in matter at extreme conditions, e.g.\ at
the occurrence of the electroweak phase transition in the early
Universe~\cite{Vachaspati:1991nm,Grasso:2000wj} or in the deep interior of
compact stellar objects like
magnetars~\cite{Duncan:1992hi,Kouveliotou:1998ze}. Moreover, it has been
pointed out that values of $|eB|$ ranging from $m_\pi^2$ to $15\,m_\pi^2$
($|B|\sim 0.3$ to $5\times 10^{19}$~G) can be reached in noncentral
collisions of relativistic heavy ions at RHIC and LHC
experiments~\cite{Skokov:2009qp,Voronyuk:2011jd}. Though these large
background fields are short lived, they should be strong enough to affect
the hadronization process, offering the amazing possibility of recreating a
highly magnetized QCD medium in the lab.

{}From the theoretical point of view, the study of strong interactions in
the presence of a large magnetic field includes several interesting
phenomena, such as the chiral magnetic
effect~\cite{Kharzeev:2007jp,Fukushima:2008xe,Kharzeev:2015znc}, which
entails the generation of an electric current induced by chirality
imbalance, and the so-called magnetic
catalysis~\cite{Klevansky:1989vi,Gusynin:1995nb} and inverse magnetic
catalysis~\cite{Bali:2011qj,Bali:2012zg}, which refer to the effect of the
magnetic field on the size of chiral quark-antiquark condensates and on the
restoration of chiral symmetry. Yet another interesting issue is the
possible existence of a phase transition of the cold vacuum into an
electromagnetic superconducting state. For a sufficiently large external
magnetic field, this transition would be induced by the emergence of
quark-antiquark vector condensates that carry the quantum numbers of
electrically charged $\rho$ mesons~\cite{Chernodub:2010qx,Chernodub:2011mc}.
The presence of such a superconducting (anisotropic and inhomogeneous) QCD
vacuum state has been discussed in the past few years and still remains as
an open problem~\cite{Braguta:2011hq,Hidaka:2012mz,Li:2013aa,Liu:2014uwa}.

It is clear that the study of the properties of light hadrons, in particular
$\pi$ and $\rho$ mesons, comes up as a crucial task towards the
understanding of the above mentioned phenomena. This represents a nontrivial
problem, since first-principle theoretical calculations require to deal in
general with QCD in a low energy nonperturbative regime. Therefore, the
corresponding theoretical analyses have been carried out using a variety of
effective models for strong interactions. The effect of intense external
magnetic fields on $\pi$ meson properties has been studied e.g.~in the
framework of Nambu-Jona-Lasinio (NJL)-like
models~\cite{Fayazbakhsh:2013cha,Fayazbakhsh:2012vr,Avancini:2015ady,Zhang:2016qrl,
Avancini:2016fgq,Mao:2017wmq,GomezDumm:2017jij,Wang:2017vtn,Liu:2018zag,
Coppola:2018vkw,Mao:2018dqe,Avancini:2018svs,Coppola:2019uyr,Cao:2019res,Sheng:2021evj,
Avancini:2021pmi}, quark-meson models~\cite{Kamikado:2013pya,Ayala:2018zat},
chiral perturbation theory
(ChPT)~\cite{Andersen:2012zc,Agasian:2001ym,Colucci:2013zoa}, path integral
Hamiltonians~\cite{Orlovsky:2013gha,Andreichikov:2016ayj}, effective chiral
confinement Lagrangians~\cite{Simonov:2015xta,Andreichikov:2018wrc} and QCD
sum rules~\cite{Dominguez:2018njv}. In addition, several results for the
$\pi$ meson spectrum in the presence of background magnetic fields have been
obtained from lattice QCD (LQCD)
calculations~\cite{Bali:2011qj,Luschevskaya:2015bea,Luschevskaya:2014lga,Brandt:2015hnz,Bali:2017ian,Ding:2020hxw}.
 Regarding the $\rho$ meson sector, studies of magnetized $\rho$ meson
masses in the framework of effective models and LQCD can be found in
Refs.~\cite{Chernodub:2011mc,Andreichikov:2016ayj,Zhang:2016qrl,Liu:2018zag,
Cao:2019res,Kawaguchi:2015gpt,Ghosh:2016evc,Ghosh:2020qvg,Avancini:2022qcp}
and
Refs.~\cite{Luschevskaya:2012xd,Luschevskaya:2015bea,Luschevskaya:2014lga,Luschevskaya:2016epp,
Bali:2017ian}, respectively.

In this work we study the mass spectrum of light neutral pseudoscalar and
vector mesons in the presence of an external uniform magnetic field $\vec
B$, considering a two-flavor NJL-like
model~\cite{Vogl:1991qt,Klevansky:1992qe,Hatsuda:1994pi}. In general, in
this type of model the calculations involving quark loops for nonzero $B$
include the so-called Schwinger phases~\cite{Schwinger:1951nm}, which are
responsible for the breakdown of translational invariance of quark
propagators. However, in the particular case of neutral mesons these phases
cancel out, and one is free to take the usual momentum basis to diagonalize
the corresponding polarization
functions~\cite{Fayazbakhsh:2013cha,Fayazbakhsh:2012vr,
Avancini:2015ady,Avancini:2016fgq,Mao:2017wmq}. One also has to care about
the regularization procedure, since the presence of the external field can
lead to spurious results, such as unphysical oscillations  of various
observables~\cite{Allen:2015paa,Avancini:2019wed}. We consider here a
magnetic field independent regularization (MFIR)
method~\cite{Menezes:2008qt,Avancini:2015ady,Avancini:2016fgq,Coppola:2018vkw},
which has been shown to be free from these effects and reduces the
dependence of the results on model parameters. In addition, in our work we
consider two mixing effects that have been mostly neglected in previous
analyses. The first one is flavor mixing in the spin zero sector; while we
restrict to a two-flavor model (keeping a reduced number of free parameters,
and assuming that strangeness does not play an essential role), we consider
quark-antiquark interactions both in $I=1$ and $I=0$ scalar and pseudoscalar
channels, introducing a 't Hooft-like effective
interaction~\cite{tHooft:1986ooh}. The second one is the mixing between
pseudoscalar and vector mesons, which arises naturally in the context of the
NJL model. These mixing contributions are usually forbidden by isospin and
angular momentum conservation, but they arise (and may become important) in
the presence of the external magnetic field. In fact, our analysis shows
that $\pi^0$~-~$\eta$~-~$\rho^0$~-~$\omega$ mixing has a substantial effect
on the $B$ dependence of the lowest mass state. As a additional ingredient,
we consider the case of $B$-dependent effective coupling constants; this
possibility
---inspired by the magnetic screening of the strong coupling constant
occurring for large $B$~\cite{Miransky:2002rp}--- has been previously
explored in effective
models~\cite{Ayala:2014iba,Farias:2014eca,Ferreira:2014kpa,Endrodi:2019whh,
Sheng:2021evj} in order to reproduce the inverse magnetic catalysis effect
observed at finite temperature in LQCD calculations.

In the case of the neutral vector mesons, we consider both states with
quantum numbers $S_z=0$ and $S_z = \pm 1$, where $S_z$ is the spin
projection in the direction of the magnetic field (it is worth noticing that
only $S_z = 0$ states can mix with pseudoscalar states). Most LQCD results
and effective model calculations agree in the finding that the masses of
$S_z= \pm 1$ states get monotonically enhanced with the magnetic field,
while results for $S_z = 0$ mesons are still not
conclusive~\cite{Luschevskaya:2012xd,Luschevskaya:2014lga,
Luschevskaya:2015bea,Andreichikov:2016ayj,Bali:2017ian,Liu:2018zag,Avancini:2022qcp}.
In our framework, which lacks a description of confinement, for large
magnetic fields the masses of some of the $S_z=0$ states are found to grow
beyond the $q\bar q$ pair production threshold; therefore our results in
this region should be taken just as qualitative ones.

The paper is organized as follows. In Sec.~II we introduce the theoretical
formalism used to obtain neutral pseudoscalar and vector meson masses. Then,
in Sec.~III we present and discuss our numerical results, while in Sec.~IV
we provide a summary of our work, together with our main conclusions. We
also include Appendixes A, B and C to provide some technical details of our
calculations.
\vfill

\section{Theoretical formalism}

\subsection{Effective Lagrangian and mean field properties}
\label{sect2a}

Let us start by considering the Euclidean Lagrangian density for an extended NJL
two-flavor model in the presence of an electromagnetic field. We have
\begin{eqnarray}
{\cal L} & = & \bar \psi(x) \left(- i\, \rlap/\!D + m_c \right) \psi(x) - \gcoups
\sum_{a=0}^3 \Big[ \left( \bar\psi(x) \taua \psi(x) \right)^2 + \left( \bar\psi(x) \, i\gamma_{5} \taua \psi(x)\right)^2 \Big] \nonumber \\
& & - \,\gcoupr \left( \bar\psi(x) \,
\gamma_\mu\vec{\tau}\,\psi(x)\right)^2 - \gcoupw  \left(
\bar\psi(x) \, \gamma_\mu\,\psi(x)\right)^2
+ \, 2 g_d \, \left(d_+ + d_-\right)\ ,
\label{lagrangian}
\end{eqnarray}
where $\psi = (u\ d)^T$, $\taua = (\mathbbm{1} , \vec \tau)$,
$\vec \tau$ being the usual Pauli-matrix vector, and $m_c$ is the
current quark mass, which is assumed to be equal for $u$ and $d$
quarks. The model includes isoscalar and isovector vector
couplings, and also a 't Hooft-like flavor-mixing term where we
have defined $d_\pm = {\rm det}[\bar \psi(x) (1 \pm \gamma_5)
\psi(x)]$. The interaction between the fermions and the
electromagnetic field ${\cal A}_\mu$ is driven by the covariant
derivative
\begin{equation}
D_\mu\ = \ \partial_{\mu}-i\,\hat Q \mathcal{A}_{\mu}\ ,
\label{covdev}
\end{equation}
where $\hat Q=\mbox{diag}(Q_u,Q_d)$, with $Q_u=2e/3$ and $Q_d = -e/3$, $e$
being the proton electric charge. We consider the particular case in which
one has a homogenous stationary magnetic field $\vec B$ orientated along the
3, or $z$, axis. Then, choosing the Landau gauge, we have $\mathcal{A}_\mu =
B\, x_1\, \delta_{\mu 2}$.

Since we are interested in studying meson properties, it is
convenient to bosonize the fermionic theory, introducing scalar,
pseudoscalar and vector fields $\sigmaa(x)$, $\pia(x)$ and
$\rhoamu(x)$, with $a=0,1,2,3$, and integrating out the fermion
fields. The bosonized
Euclidean action can be written as 
\begin{eqnarray}
S_{\mathrm{bos}} & = & -\ln\det\mathcal{D} +\frac{1}{4\gcoupt}
\int d^{4}x\ \Big[\sigmac(x)\, \sigmac(x)+
\vec{\pi}(x)\cdot\vec{\pi}(x)\Big]
\nonumber \\
& & +\, \frac{1}{4\gcoupt(1-2 \cc)} \int d^{4}x\ \Big[\vec
\sigma(x)\cdot\vec \sigma(x)+ \pic(x)\, \pic(x)\Big]
\nonumber \\
& & + \, \frac{1}{4\gcoupr} \int d^{4}x\ \vec\rho_\mu (x)\cdot\vec\rho_\mu (x)
+ \frac{1}{4\gcoupw} \int d^{4}x\ \rhocmu (x) \, \rhocmu (x) \ ,
\label{sbos}
\end{eqnarray}
with
\begin{equation}
\mathcal{D}_{x,x'} \ = \ \delta^{(4)}(x-x')\,\big[-i\,\rlap/\!D + m_0 +
\taua \left( \sigmaa(x) + i\,\gamma_5\, \pia(x) +
\gamma_\mu \, \rhoamu(x) \right)\big] \ ,
\label{dxx}
\end{equation}
where a direct product to an identity matrix in color space is understood.
Note that for convenience we have introduced the combinations
\begin{eqnarray}
\gcoupt = \gcoups + g_d \ ,  \qquad \qquad \cc = g_d/(\gcoups + g_d)\ ,
\end{eqnarray}
so that the flavor mixing in the scalar-pseudoscalar sector is regulated by
the constant $\cc$. For $\cc=0$ quark flavors $u$ and $d$ get decoupled,
while for $\cc=0.5$ one has maximum flavor mixing, as in the case of
the standard version of the NJL model.

We proceed by expanding the bosonized action in powers of the fluctuations
of the bosonic fields around the corresponding mean field (MF) values. We
assume that the fields $\sigma_a(x)$ have nontrivial translational invariant
MF values given by $\tau_a \bar\sigma_a = \mbox{diag}(\bar \sigma_u,\bar
\sigma_d)$, while vacuum expectation values of other bosonic fields are
zero; thus, we write
\begin{equation}
\mathcal{D}_{x,x'} \ = \ \mathcal{D}^{\mbox{\tiny MF}}_{x,x'} + \delta\mathcal{D}_{x,x'}\ .
\label{dxxp}
\end{equation}
The MF piece is diagonal in flavor space. One has
\begin{equation}
\mathcal{D}^\mf_{x,x'} \ = \ {\rm diag}\big(\mathcal{D}^{\mf,\,u}_{x,x'}\, ,\,
\mathcal{D}^{\mf,\,d}_{x,x'}\big)\ ,
\end{equation}
with
\begin{equation}
\mathcal{D}^{\mf,\,f}_{x,x'} \ = \ \delta^{(4)}(x-x') \left( - i \rlap/\partial
- Q_f \, B \, x_1 \, \gamma_2 + M_f \right) \ ,
\end{equation}
where $M_f=m_c+\bar\sigma_f$ is the quark effective mass for each flavor
$f$.

The MF action per unit volume is given by
\begin{equation}
\frac{S^{\mbox{\tiny MF}}_{\mathrm{bos}}}{V^{(4)}} \ = \
\frac{ (1-\cc) (\bar \sigma_u^2+\bar \sigma_d^2) -2\, \cc\, \bar \sigma_u \bar \sigma_d }{8 \gcoupt (1-2\, \cc)}
- \frac{N_c}{V^{(4)}} \sum_{f=u,d} \int d^4x \, d^4x' \
\trmin_D\, \ln \left(\mathcal{S}^{\mbox{\tiny MF},\,f}_{x,x'}\right)^{-1} \ ,
\label{seff}
\end{equation}
where $\trmin_D$ stands for the trace in Dirac space, and
$\mathcal{S}^{\mf,\,f}_{x,x'} = \big( \mathcal{D}^{\mf,\,f}_{x,x'} \big)^{-1}$
is the MF quark propagator in the presence of the magnetic field. As is well
known, the explicit form of the propagators can be written in different
ways~\cite{Andersen:2014xxa,Miransky:2015ava}. For convenience we take the
form in which $\mathcal{S}^{\mf,\,f}_{x,x'}$ is given by a product of a phase
factor and a translational invariant function, namely
\begin{equation}
\mathcal{S}^{\mf,\,f}_{x,x'} \ = \ e^{i\Phi_f(x,x')}\,\int_p e^{ip\, (x-x')}\,
\tilde S_p^f \ ,
\label{sfx}
\end{equation}
where $\Phi_f(x,x')=q_f B (x_1+x_1')(x_2-x_2')/2$ is the
so-called Schwinger phase. We have introduced here the shorthand notation
\begin{equation}
\int_{p}\ \equiv \ \int \dfrac{d^4 p}{(2\pi)^4}\ .
\label{notation1}
\end{equation}
Now $\tilde S_p^f$ can be expressed in the Schwinger
form~\cite{Andersen:2014xxa,Miransky:2015ava}
\begin{eqnarray}
\tilde S_p^f & = & \int_0^\infty d\tau\
\exp\!\bigg[-\tau\Big(M_f^2+p_\parallel^2+p_\perp^2\,\dfrac{\tanh(\tau B_f)}{\tau B_f} - i \epsilon\Big) \bigg] \nonumber\\
& & \times \, \bigg\{\big(M_f-p_\parallel \cdot \gamma_\parallel\big)\,\big[1+i s_f \, \gamma_1 \gamma_2 \, \tanh(\tau B_f)\big] -
\dfrac{p_\perp \cdot \gamma_\perp}{\cosh^2 (\tau B_f)}  \bigg\}\ ,
\label{sfp_schw}
\end{eqnarray}
where we have used the following definitions. The perpendicular and parallel
gamma matrices are collected in vectors $\gamma_\perp = (\gamma_1,\gamma_2)$
and $\gamma_\parallel = (\gamma_3,\gamma_4)$, and, similarly, we have
defined $p_\perp = (p_1,p_2)$ and $p_\parallel = (p_3,p_4)$. Note that
we are working in Euclidean space, where $\{ \gamma_\mu, \gamma_\nu \} = -2
\delta_{\mu\nu}$. Other definitions in Eq.~(\ref{sfp_schw}) are $s_f = {\rm
sign} (Q_f B)$ and $B_f=|Q_fB|$. The limit $\epsilon \rightarrow 0$  is
implicitly understood.

The integral in Eq.~\eqref{sfp_schw} is divergent and has to be
properly regularized. As stated in the Introduction, we use here
the magnetic field independent regularization (MFIR) scheme: for a
given unregularized quantity, the corresponding (divergent) $B \to
0$ limit is subtracted and then it is added in a regularized form.
Thus, the quantities can be separated into a (finite) ``$B=0$''
part and a ``magnetic'' piece. Notice that, in general, the
``$B=0$" part still depends implicitly on $B$ (e.g.\ through the
values of the dressed quark masses $M_f$), hence it should not be
confused with the value of the studied quantity at vanishing
external field. To deal with the divergent ``$B=0$'' terms we use
here a 3D cutoff regularization scheme. In the case of the
quark-antiquark condensates $\phi_f \equiv \langle \bar \psi_f
\,\psi_f \rangle$, $f=u,d$, we obtain
\begin{equation}
\phi_f^{\rm reg} \ = \ \phi_f^{0,\,\rm reg} \, + \, \phi_f^{\rm mag} \ ,
\end{equation}
where
\begin{equation}
\phi_f^{0,\rm reg} = -N_c\, M_f \, I_{1f}\ , \qquad\qquad
\phi_f^{\rm mag} = -N_c\, M_f \,I_{1f}^{\rm mag}\ .
\label{phif}
\end{equation}
The expression of $I_{1f}$ for the 3D cutoff regularization is given by
Eq.~\eqref{I1freg} of App.\ A~\cite{Klevansky:1992qe}, while the $B$-dependent
function $I_{1f}^{\mbox{\tiny mag}}$
reads~\cite{Allen:2015paa,Klevansky:1989vi}
\begin{equation}
I_{1f}^{\rm mag} = \dfrac{B_f}{2\pi^2} \left[ \ln \Gamma(x_f) -
\left(x_f - \dfrac{1}{2}\right) \ln x_f + x_f - \dfrac{\ln{2\pi}}{2} \right]
\, , \label{i1}
\end{equation}
where $x_f=M_f^2/(2B_f)$. The corresponding gap equations, obtained from
$\partial S^{\mbox{\tiny MF}}_\mathrm{bos} / \partial \bar{\sigma}_f =0$,
can be written as
\begin{eqnarray}
 M_{u} & = & m_c - 4 \gcoupt \left[ (1-\cc) \, \phi_u^{\rm reg} + \cc \, \phi_d^{\rm reg}\right] \ , \nonumber\\
 M_{d} & = & m_c - 4 \gcoupt \left[ (1-\cc) \,  \phi_d^{\rm reg} + \cc \, \phi_u^{\rm reg}\right]\ .
\label{gapeqs}
\end{eqnarray}
As anticipated, for $\alpha=0$ these equations get decoupled. For
$\alpha=0.5$ the right hand sides become identical, thus one has in that
case $M_u=M_d$.

\subsection{Neutral meson system}
\label{neutral}

As expected from charge conservation, it is easy to see that the
contributions to the bosonic action that are quadratic in the fluctuations
of charged and neutral mesons decouple from each other. In this work we
concentrate on the neutral meson sector. For notational convenience we will
denote {\rm isospin states} by $M=\sigmaa,\pia,\rhoamu$, with $a=0,3$. Here
$\sigmac$, $\pic$ and $\rhoc$ correspond to the isoscalar states
$\sigma$, $\eta$ and $\omega$, while $\sigmat$, $\pit$ and $\rhot$
stand for the neutral components of the isovector triplets $\vec a_0$, $\vec
\pi$ and $\vec \rho$, respectively. Thus, the corresponding quadratic piece
of the bosonized action can be written as
\begin{equation}
S^{\rm quad,\, neutral}_{\mathrm{bos}} \ = \ \frac{1}{2} \int d^4x\; d^4 x'
\sum_{M,M'}\delta M(x)  \ {\gmatrix}_{MM'}(x,x') \  \delta M'(x')\ .
\end{equation}
The functions ${\gmatrix}_{MM'}(x,x')$ can be separated in two terms, namely
\begin{equation}
{\gmatrix}_{MM'}(x,x') \ = \  \frac{1}{2 g_M}\ \delta_{MM'} \ \delta^{(4)}(x-x')+
{\jmatrix}_{MM'}(x,x') \ ,
\end{equation}
where $\delta_{MM'}$ is an obvious generalization of the Kronecker $\delta$,
and the constants $g_M$ are given by
\begin{equation}
g_M \ = \
\left\{%
\begin{array}{cc}
  \gcoupt                &\ \  \ \ \mbox{for}\ \ M= \sigmac, \pit \\
  \gcoupt (1-2\cc)         &\ \  \ \ \mbox{for}\ \ M= \sigmat, \pic \\
  \gcoupr                &\mbox{for}\ \ M= \rhotmu \\
  \gcoupw                &\mbox{for}\ \ M= \rhocmu \\
\end{array}%
\right. \qquad .
\label{valgm}
\end{equation}
The polarization functions ${\jmatrix}_{MM'}(x,x')$ can be separated into
$u$ and $d$ quark pieces,
\begin{equation}
{\jmatrix}_{M M'} (x,x') \ = \ {\cal F}^u_{M M'}(x',x)\, + \, \varepsilon_M\,
\varepsilon_{M'} \, {\cal F}^d_{M M'}(x',x)\ .
\label{jotas}
\end{equation}
Here $\varepsilon_M = 1$ for the isoscalars $M= \sigmac, \pic ,
\rhocmu$ and $\varepsilon_M = -1$ for $M= \sigmat, \pit , \rhotmu$,
while the functions ${\cal F}^f_{M M'}(x',x)$ are found to be
\begin{eqnarray}
{\cal F}^f_{M M'}(x',x) \ = \
N_c \ \trmin_D \bigg[ \mathcal{S}^{\mf,\,f}_{x,x'}\, \Gamma^{M'}
\mathcal{S}^{\mf,\,f}_{x',x}\, \Gamma^{M}\, \bigg]\ ,
\end{eqnarray}
with
\begin{equation}
\Gamma^M \ = \
\left\{%
\begin{array}{cc}
  1 &  \mbox{for}\ \ M= \sigmac, \sigmat \\
  i \gamma_5 & \mbox{for}\ \ M= \pic, \pit \\
  \gamma_\mu & \mbox{for}\ \ M= \rhocmu, \rhotmu
\end{array}%
\right. \qquad .
\end{equation}

As stated, since we are dealing with neutral mesons, the contributions of
Schwinger phases associated with the quark propagators in Eq.~(\ref{sfx})
cancel out, and the polarization functions depend only on the difference
$x-x'$, i.e., they are translationally invariant. After a Fourier
transformation, the conservation of momentum implies that the polarization
functions turn out to be diagonal in the momentum basis. Thus, in this basis
the neutral meson contribution to the quadratic action can be written as
\begin{eqnarray}
S^{\rm quad,\,neutral}_{\mathrm{bos}} \ = \ \frac{1}{2}
\sum_{M,M'} \int_q \  \delta M(-q) \ {\gnomatrix}_{MM'}(q) \ \delta M'(q)\ .
\label{eqnneutral}
\end{eqnarray}
Now we have
\begin{equation}
{\gnomatrix}_{MM'}(q) \ = \ \frac{1}{2 g_M}\;\delta_{MM'} \,+\,
{\jnomatrix}_{MM'}(q)\ ,
\label{gmm}
\end{equation}
and the associated polarization functions are given by
\begin{equation}
{\jnomatrix}_{M M'} (q) \ = \ F^u_{M M'}(q) + \varepsilon_M
\varepsilon_{M'} \, F^d_{M M'}(q)  \ .
\label{jotas}
\end{equation}
The functions $F^f_{M M'}(q)$ read
\begin{equation}
F^f_{M M'}(q) \ = \ N_c \, \int_p \ \trmin_D \left[ \tilde S_{p_+}^f  \Gamma^{M'} \tilde S_{p_-}^f  \Gamma^{M} \right]\ ,
\end{equation}
where we have defined $p_\pm = p \pm q/2$, and the quark propagators $\tilde
S_p^f$ in the presence of the magnetic field have been given in
Eq.~(\ref{sfp_schw}).

It is relatively easy to see that the functions
${\jnomatrix}_{\sigmaa\pib}(q)$ are zero for either $a$ or $b$ equal to 0
or 3. However, the remaining polarization functions do not vanish in
general. Since we are interested in the determination of meson masses, we
consider here the particular case in which mesons are at rest, i.e.\ we take
$\vec q =0$, $q_4^2=-m^2$, where $m$ stands for the corresponding meson
mass. In that situation the nondiagonal polarization functions that mix the
neutral scalar and vector mesons also vanish, i.e.\ for $a,b=0,3$ one has
$\hat {\jnomatrix}_{\sigmaa\rhobmu}=0$, where the notation $\hat
{\jnomatrix}$ indicates that the polarization function is evaluated at the
meson rest frame. In this way, the scalar meson sector gets decoupled at
this level; we will not take into account these mesons in what follows. It
can also be shown that $\hat {\jnomatrix}_{\rhoamu\rhobnu}$, with
$a,b=0,3$, vanish for $\mu \neq \nu$, while the functions $\hat
{\jnomatrix}_{\pia\rhobmu}$, with $a,b=0,3$, turn out to be proportional
to $\delta_{\mu 3}$.

It is found that all nonvanishing polarization functions are in
general divergent. As done at the MF level, we consider the
magnetic field independent regularization scheme, in which we
subtract the corresponding ``$B=0$'' contributions and then we add
them in a regularized form. Thus, for a generic polarization
function $\hat {\jnomatrix}_{M M'}$ we have
\begin{equation}
\hat {\jnomatrix}^{\rm reg}_{M M'} \ = \ \hat {\jnomatrix}^{0,{\rm reg}}_{M M'} \, + \,
\hat {\jnomatrix}^{\rm mag}_{M M'}\ .
\label{jregjmag}
\end{equation}
The regularized ``$B=0$'' pieces $\hat {\jnomatrix}^{0,{\rm
reg}}_{M M'}$ are given in App.\ A;  it is easy to see that all
nondiagonal polarization functions $\hat {\jnomatrix}^{0,\,{\rm
reg}}_{MM'}$, $M \neq M'$, are equal to zero. In the case of the
``magnetic'' contributions $\hat {\jnomatrix}^{\rm mag}_{M M'}$,
after a rather long calculation it is found that they can be
expressed in the form given by Eq.~(\ref{jotas}), viz.
\begin{equation}
\hat {\jnomatrix}^{\rm mag}_{M M'} \ = \ \hat F^{u,\,{\rm mag}}_{M M'} + \varepsilon_M
\varepsilon_{M'} \, \hat F^{d,\,{\rm mag}}_{M M'}  \ ,
\end{equation}
where the functions $\hat F^{f,\,{\rm mag}}_{M M'}$ are given by
\begin{eqnarray}
\hat {F}_{\pi^a\pi^b}^{f,\,{\rm mag}}
&=&
- N_c \left[ I_{1f}^{\rm mag} - m^2 I^{\rm mag}_{2f}(-m^2) \right]\ ,
\nonumber
\\[2.mm]
\hat {F}_{\pi^a\rho_\mu^b}^{f,\,{\rm mag}}&=&-\hat {F}_{\rho_\mu^a\pi^b}^{f,\,{\rm mag}}
= i N_c\, I^{\rm mag}_{3f}(-m^2)\, \delta_{\mu 3}\ ,
\nonumber
\\[2.mm]
\hat {F}_{\rho_\mu^a\rho_\nu^b}^{f,\,{\rm mag}}
&=&
N_c \left[ I^{\rm mag}_{4f}(-m^2)
\, \mathbbm{1}^\perp_{\mu\nu} + m^2  I^{\rm mag}_{5f}(-m^2)
\, \delta_{\mu 3} \delta_{\nu 3}\right]\ ,
\label{cfmag}
\end{eqnarray}
with $\mathbbm{1}^\perp = \mbox{diag}(1,1,0,0)$. The expression for
$I_{1f}^{\rm mag}$ has been given in Eq.~(\ref{i1}), whereas the integrals
$I^{\rm mag}_{nf}$ for $n=2,\dots ,5$ read
\begin{eqnarray}
I^{\rm mag}_{2f}(-m^2) &=& \frac{1}{8 \pi^2} \int_0^1 dv \left[ \psi(\bar
x_f) + \frac{1}{2 \bar x_f} - \ln \bar x_f \right]\ ,
\nonumber \\
I^{\rm mag}_{3f}(-m^2) &=& \frac{s_f M_f B_f}{\pi^2 m} \int_0^1 dv
\, \left( v^2 + 4M_f^2/m^2 -1 \right)^{-1}\ ,
\nonumber \\
I^{\rm mag}_{4f}(-m^2) &=& -\, I^{\rm mag}_{1f}  -
\frac{m^2}{16 \pi^2} \Bigg[ \int_0^1 dv \ (v^2+ \gamma ) \ln \bar
x_f
\nonumber \\
& & -\, \frac{1}{2} \sum_{s=\pm 1}\int_0^1 dv \ ( v^2 + s v/\lambda + \gamma)
\ \psi(\bar x_f + (1+s v)/2) \Bigg]\ ,
\nonumber \\
I^{\rm mag}_{5f}(-m^2) &=& \frac{1}{8 \pi^2} \int_0^1 dv \ (1 - v^2)
\left[ \psi(\bar x_f) + \frac{1}{2 \bar x_f} - \ln \bar x_f \right]\ ,
\label{inf}
\end{eqnarray}
where $\lambda = m^2/(4 B_f)$, $\gamma = 1+4M_f^2/m^2$ and $\bar x_f =
\left[ M_f^2 - (1-v^2) m^2/4 \right] /(2 B_f)$. For $m < 2 M_f$ these
integrals are well defined. In fact, in the case of $I^{\rm mag}_{3f}(-m^2)$
one can even get the analytic result
\begin{equation}
I^{\rm mag}_{3f}(-m^2) \ = \ \frac{s_f\, B_f}{2\pi^2\, \sqrt{1-m^2/(4
M_f^2)}}\; \arctan\Bigg(\frac{m}{2M_f\sqrt{
1-m^2/(4M_f^2)}}\Bigg)\ ,\qquad m < 2M_f\ .
\end{equation}
In the case of $I^{\rm mag}_{4f}(-m^2)$, it is worth noticing that in the
limit $m^2\to 0$ the second term on the r.h.s.\ of the corresponding
expression in Eqs.~(\ref{inf}) is found to be equal to $I^{\rm mag}_{1f}$.
Thus, in this limit one has $I^{\rm mag}_{4f}\to 0$, as it is required in
order to avoid a nonzero contribution to the photon mass coming from the
``magnetic piece'' of the polarization function. On the other hand, for
$m\geq 2M_f$ (i.e., beyond the $q\bar q$ production threshold) the integrals
are divergent. To obtain finite results we perform in this case analytic
extensions. The corresponding expressions, as well as some technical
details, are given in App.~B.

The vector fields $\rhocmu$ and $\rhotmu$ can be written in a
polarization vector basis. Since we assume that the mesons are at rest, we
can choose polarization vectors $\epsilon_\mu^{(S_z)}$ associated to spin
projections $S_z=0, \pm 1$, namely
\begin{equation}
\epsilon_\mu^{(0)} = \left(
  0 ,
  0 ,
  1 ,
 0
\right)
\ , \qquad
\epsilon_\mu^{(1)} = \frac{1}{\sqrt 2} \left(
  1 ,
  i ,
  0 ,
 0
\right)
\ , \qquad
\epsilon_\mu^{(-1)} = \frac{1}{\sqrt 2} \left(
  1 ,
  -i ,
  0 ,
 0
\right)
\ .
\end{equation}
It is convenient to distinguish between states $\rhoapar$ and $\rhoaperp$,
with $a=0,3$, corresponding to polarization vectors parallel and
perpendicular to the magnetic field (spin projections $S_z=0$ and $S_z=\pm
1$), respectively. From Eqs.~(\ref{cfmag}) it is seen that for nonzero $B$
pseudoscalar mesons get coupled only to neutral vector mesons with spin
projection $S_z=0$ (in fact, this is expected from the invariance under
rotations around the direction of $\vec B$). In this way, taking into
account Eq.~(\ref{gmm}) one can define a $4\times 4$ matrix $G_\parallel$
with elements $G_{M M'}$, where $M,M'=\pic,\pit,\rhocpar,\rhotpar$, whereas
for $S_z = \pm 1$ states one gets two identical $2\times 2$ matrices
$G_{\perp}$ with elements $G_{M M'}$, where $M,M'=\rhocperp,\rhotperp$. The
pole masses of $S_z=0$ physical mesons, $m^{(k)}_\parallel$ (with
$k=1,\dots,4$), will be given by the solutions of
\begin{equation}
\mbox{det}\,G_\parallel \ = \ 0 \ ,
\label{detgpar}
\end{equation}
while those of the $S_z = \pm 1$ vector mesons, $m^{(k)}_{\perp}$ (with
$k=1,2$), can be obtained from
\begin{equation}
\mbox{det}\, G_{\perp} \ = \ 0 \ .
\label{detgperp}
\end{equation}
Once the masses are determined, the spin-isospin composition of the physical
meson states $|k\rangle$ is given by the corresponding eigenvectors
$c^{(k)}$. Thus, one has
\begin{equation}
\begin{array}{ll}
|k\rangle = c^{(k)}_{\pic} \, |\pic\rangle  + c^{(k)}_{\pit} \, |\pit\rangle  +
i\, c^{(k)}_{\rhocpar}\, |\rhocpar\rangle + i\, c^{(k)}_{\rhotpar} \,
|\rhotpar\rangle\ , \
k=1,\dots ,4 & \qquad {\rm for} \ S_z=0\ {\rm states}\ ,
\\
|k\rangle = c^{(k)}_{\rhotperp}\, |\rhocperp\rangle + c^{(k)}_{\rhotperp} \, |\rhotperp\rangle
\ , \ k=1,2  & \qquad {\rm for} \ S_z=\pm 1\ {\rm states}\ .
\end{array}
\label{decompara}
\end{equation}

It is also useful to consider the flavor basis $\pif$, $\rhof$, where
$f=u,d$. Isospin states can be written in terms of flavor states using the
relations
\begin{eqnarray}
& & | \pic \rangle = \frac{1}{\sqrt2}\left(| \piu \rangle + | \pid \rangle\right)\ , \qquad \qquad
| \pit \rangle  = \frac{1}{\sqrt2}\left(| \piu \rangle - | \pid \rangle\right) \ , \nonumber \\
& & | \rhocpar \rangle = \frac{1}{\sqrt2}\left(| \rhoupar \rangle + | \rhodpar \rangle\right) \ ,
\qquad\ \
| \rhotpar \rangle  = \frac{1}{\sqrt2}\left(| \rhoupar \rangle - | \rhodpar \rangle\right)\ .
\label{udbasis}
\end{eqnarray}
In the $S_z = \pm 1$ sector, where there is no mixing between pseudoscalar
and vector mesons, the states $|\rhouperp\rangle$ and
$|\rhodperp\rangle$ turn out to be the mass eigenstates that diagonalize
$G_{\perp}$. This can be easily understood noticing that the external
magnetic field distinguishes between quarks that carry different electric
charges, and this is what breaks the $u$-$d$ flavor degeneracy. In the
flavor basis one has $G_\perp = {\rm diag}(G_{u\perp}, G_{d\perp})$, where
\begin{equation}
G_{f\perp}(-m^2) \ = \ \frac{1}{2g_v}\, + \, \frac{4N_c}{3}\Big[
(2M_f^2+m^2)\,I_{2f}(-m^2)-2M_f^2\, I_{2f}(0)\Big]\, + \, 2N_c\,
I^{\rm mag}_{4f}(-m^2)\ ,
\label{gfperp}
\end{equation}
where the expression for $I_{2f}(q^2)$ can be found in App.~A, and $I^{\rm
mag}_{4f}(-m^2)$ has been given in Eqs.~(\ref{inf}). A similar situation
occurs in the $S_z = 0$ sector if one has $\alpha=0$. In this particular
case there is no flavor mixing either in the pseudoscalar or vector meson
sectors, hence the $4\times 4$ matrix $G_\parallel$ can be written as a
direct sum of $2\times 2$ flavor matrices $G_{u\parallel}$ and
$G_{d\parallel}$. Moreover, for a given value of $B$, the meson masses of
e.g.\ $u$-like mesons (solutions of the equation ${\rm det}\,
G_{u\parallel}=0$) can be obtained from those of $d$-like mesons for $B'=2
B$, since $|Q_u| = 2 |Q_d|$ and ${\rm det}\, G_{f\parallel}$ depends on $Q_f$
and $B$ only through the combination $B_f= |Q_f B|$ (this also holds for the
implicit dependence on $Q_f$ and $B$ through the quark effective masses
$M_f$). If one has $\alpha\neq 0$ this relation is no longer valid, and in
general $G_\parallel$ cannot be separated into flavor pieces. In fact, as we
discuss below, in the pseudoscalar sector it is seen that chiral symmetry
largely dominates over flavor symmetry; for the range of values of $B$
considered in this work, we find that even for $\alpha \ll 1$ the lightest
$S_z=0$ mass eigenstates are very close to isospin states $\pit$ and
$\pic$, instead of approximating to flavor states $\pif$.

\section{Numerical results}

\subsection{Model parametrization and mean field results}

To obtain numerical results for the dependence of meson masses on the
external magnetic field, one first has to fix the parameters of the model.
Here we take the parameter set $m_c = 5.833$~MeV, $\Lambda = 587.9$~MeV and
$g\Lambda^2 = 2.44$, which ---for vanishing external field--- lead to
effective quark masses $M_f=400$~MeV and quark-antiquark condensates
$\phi_f^{0} = (-241\ {\rm MeV})^3$, for $f=u,d$. This parametrization
properly reproduces the empirical values of the pion mass and decay constant
in vacuum, namely $m_\pi=138$~MeV and $f_\pi=92.4$~MeV. Regarding the vector
couplings, we take $g_{v_3}= 2.651/\Lambda^2$, which leads to
$m_\rho=770$~MeV at $B=0$, and $g_{v_0} = g_{v_3}$, which is consistent with
the fact that $m_\rho \simeq m_\omega$ at vanishing external field. For
these constants we use from now on the notation $g_v \equiv g_{v_0} =
g_{v_3}$. Finally, as stated in Sec.~\ref{sect2a}, the amount of flavor
mixing induced by the 't Hooft-like interaction is controlled by the
parameter $\alpha$. In this work we choose to take as a reference value
$\alpha=0.1$, since it leads (at $B=0$) to an approximate $\eta$ meson mass
$m_\eta \simeq 520$~MeV, in reasonable agreement with the physical value
$m^{\rm phys}_\eta = 548$~MeV. In fact, this mass is very sensitive to minor
changes in $\alpha$. An alternative estimate for this parameter can be
obtained from the $\eta - \eta'$ mass splitting within the 3-flavor NJL
model~\cite{Kunihiro:1989my}, which leads to $\alpha \simeq
0.2$~\cite{Frank:2003ve}. In any case, to obtain a full understanding of the
effects of flavor mixing we will also consider the values $\alpha=0$ and
$\alpha=0.5$, corresponding to the situation in which flavors are decoupled
and in which there is full flavor mixing, respectively. It is easily seen
that for $\alpha =0$ the $\pi$ and $\eta$ mesons have equal (finite) masses,
while when $\alpha$ approaches 0.5 the mass of the pion stays finite and
that of the $\eta$ meson becomes increasingly large.

\begin{figure}[tb]
     \centering{}\includegraphics[width=0.75\textwidth]{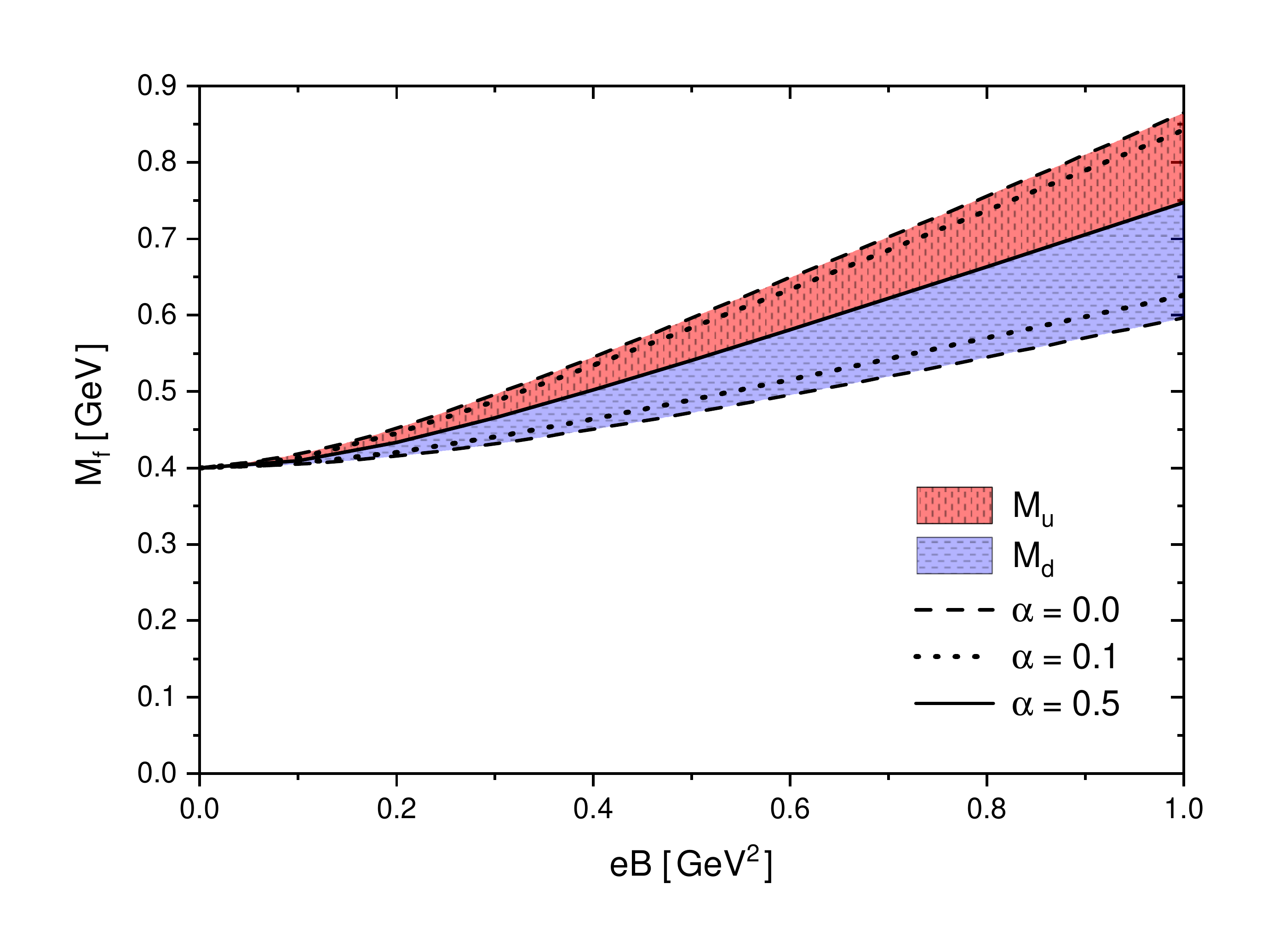}
\caption{(Color online) Effective quark masses $M_u$ (red upper band),
$M_d$ (blue lower band) as functions of $eB$. The extremes of the bands
correspond to $\alpha=0$ (dashed lines) and $\alpha=0.5$ (full
line). The dotted lines correspond to $\alpha=0.1$. }
\label{Fig1}
\end{figure}

\begin{figure}[tb]
     \centering{}\includegraphics[width=0.9\textwidth]{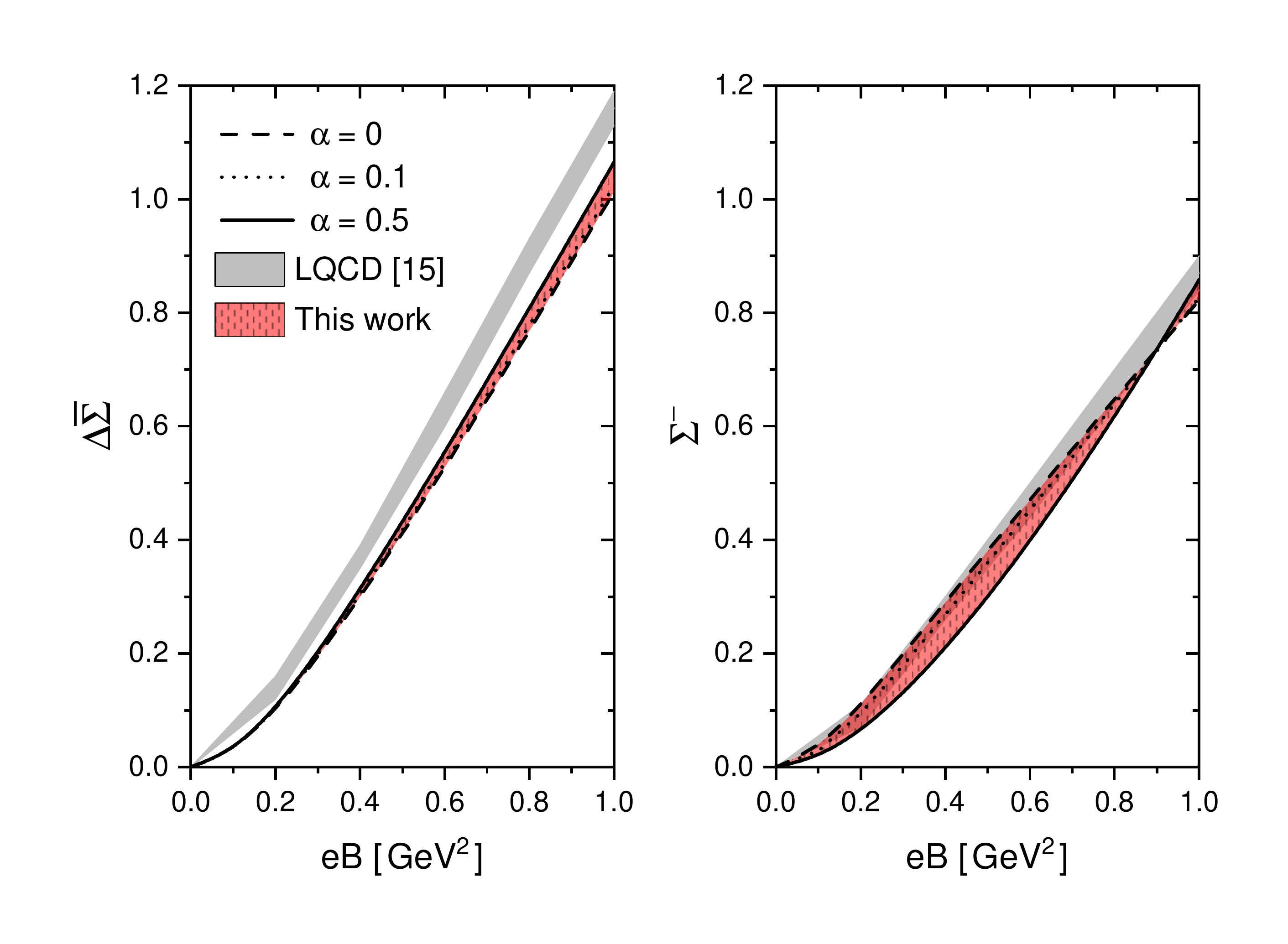}
\caption{(Color online) Normalized average condensate (left) and normalized
condensate difference (right) as functions of $eB$, for values of $\alpha$
from 0 to 1 (see text for definitions). LQCD results from
Ref.~\cite{Bali:2012zg} (gray bands) are added for comparison.}
\label{Fig2}
\end{figure}

In Fig.~\ref{Fig1} we show the numerical results obtained for the
magnetic field dependence of the dynamical quark masses $M_u$ and
$M_d$.
Both masses are found to get increased with $B$, and it is seen that for
$M_u$ ($M_d$) the slope becomes larger (smaller) as $\alpha$ decreases from
$\alpha =0.5$
---where both masses coincide--- to $\alpha=0$. Next, in
Fig.~\ref{Fig2}, we show the dependence of normalized light quark-antiquark
condensates on $B$. Following Ref.~\cite{Bali:2012zg}, we
introduce the definitions
\begin{equation}
\Delta \bar\Sigma \ = \ \dfrac{\Delta \Sigma_u +\Delta \Sigma_d
}{2} \ \ , \qquad \Sigma^- \ = \ \Delta \Sigma_u - \Delta \Sigma_d
\ ,
\end{equation}
where $\Delta\Sigma_f =  - 2\,m_c\, [\phi_f(B) - \phi_f^0]/D^4$,
$D=(135\times 86)^{1/2}$~MeV being a phenomenological normalization
constant.
In the left and right panels of Fig.~\ref{Fig2} we plot the values
of $\Delta \bar \Sigma$ and $\Sigma^-$, respectively, as functions
of $eB$. The gray bands correspond to LQCD values taken from
Ref.~\cite{Bali:2012zg}, whereas the red bands cover our results
for the range $\alpha =0$ to $\alpha = 0.5$. We observe from this
figure that the model  reproduces properly the zero-temperature
magnetic catalysis found in LQCD calculations. Moreover, it is
seen that the dependence on the flavor mixing parameter $\alpha$
is rather mild.

\subsection{Pseudoscalar and $S_z=0$ vector meson sector}
\label{sec_sz0}

\begin{figure}[h]
     \centering{}\includegraphics[width=0.75\textwidth]{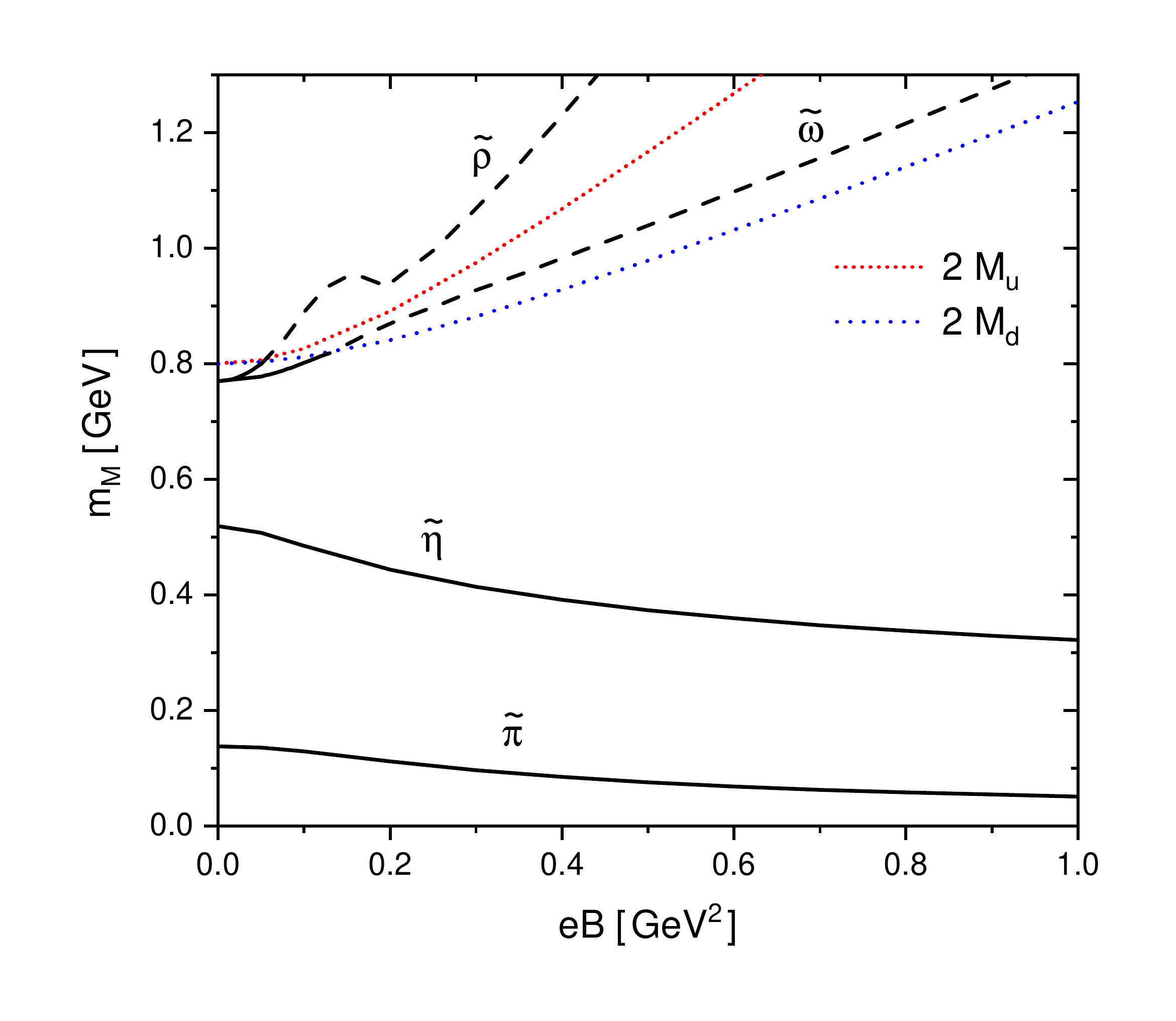}
\caption{(Color online) Masses of $S_z=0$ mesons as functions of $eB$, for
$\alpha=0.1$. The dotted lines indicate $u\bar u$ and $d\bar d$ production
thresholds.}
\label{Fig3}
\end{figure}

In this subsection we present and discuss the results associated with the
coupled system composed by neutral pseudoscalar mesons and $S_z=0$ neutral
vector mesons. As discussed in Sec.~\ref{neutral}, the corresponding masses
$m_\parallel^{(k)}$, $k=1,\dots,4$, can be obtained from
Eq.~(\ref{detgpar}). The dependence of these masses with the magnetic field
for the reference value $\alpha=0.1$ are shown in Fig.~\ref{Fig3}. As
discussed below,  the spin-isospin compositions of the associated states do
not coincide in general with those of the usual $B=0$ states $\pi^{0}$,\
$\eta$,\ $\rho^{0}$ and $\omega$. For this reason, we use for these states
the notation $\tilde M$, where in each case $M$ is the state that has the
larger weight $c_M^{(k)}$ in the spin-isospin decomposition given by
Eq.~(\ref{decompara}) (see Table~\ref{taba}).
In Fig.~\ref{Fig3} we also show the $q\bar q$ production thresholds $m=2
M_d$ and $m=2 M_u$ (dotted and short-dotted lines, respectively), beyond
which some of the matrix elements of $G_\parallel$ get absorptive parts. The
presence of these absorptive parts implies that for the states $\tilde\rho$
and $\tilde\omega$ there are certain values of the magnetic field above
which the associated particles are unstable with respect to an unphysical
decay into a $q \bar q$ pair. In fact, the existence of such decays is a
well known feature of the NJL model, even in the absence of an external
field~\cite{Vogl:1991qt,Klevansky:1992qe,Hatsuda:1994pi};  it arises as a
consequence of the lack of a confinement mechanism, which is a
characteristic of this type of model. In the presence of the magnetic field,
one also has to deal with new poles that may arise from the thresholds
related to the Landau level decomposition of the intermediate quark
propagators. As customary, we will assume that the widths associated to
these unphysical decays are small. Then, to determine the values of the
corresponding masses, we consider an extremum condition for the meson
propagators, similar to the method discussed e.g.\ in
Ref.~\cite{Bernard:1997wp}. It has to be kept in mind, however, that these
predictions for the meson masses are less reliable in comparison to those
obtained for the states lying below the quark pair production threshold, and
should be taken just as qualitative results. For this reason, in
Fig.~\ref{Fig3} we use dashed lines to plot $\tilde\rho$ and $\tilde\omega$
masses above the $2M_d$ threshold.

It is interesting at this stage to discuss the spin-isospin composition of
mass states and their variation with the external field. As mentioned at the
end of Sec.~\ref{neutral}, the magnetic field tends to separate the states
according to  the charges of the quark components. In the case $\alpha=0$,
although there is no flavor mixing, flavor degeneracy gets broken due to the
magnetic field. Therefore, mass eigenstates turn out to be separated into
particles with pure $u$ or $d$ quark content. If we use the labels $k=1,3$
and $k=2,4$ for ``$u$'' and ``$d$'' states respectively, we get [see
Eqs.~(\ref{udbasis})]
\begin{eqnarray}
\!\!\!|k\rangle \ &=& \ c^{(k)}_{\piu} | \piu \rangle  +
i\, c^{(k)}_{\rhoupar} |\rhoupar \rangle  \ = \
\frac{c^{(k)}_{\piu}}{\sqrt2} | \pic \rangle +
\frac{c^{(k)}_{\piu}}{\sqrt2} | \pit \rangle +
i\, \frac{c^{(k)}_{\rhoupar}}{\sqrt2} | \rhocpar
\rangle + i\, \frac{c^{(k)}_{\rhoupar}}{\sqrt2} |
\rhotpar \rangle\ , \ k=1,3\ ,\ \
\nonumber \\
\!\!\!|k\rangle \ &=& \ c^{(k)}_{\pid} | \pid \rangle  +
i\, c^{(k)}_{\rhodpar} |\rhodpar \rangle  \ = \
\frac{c^{(k)}_{\pid}}{\sqrt2} | \pic \rangle -
\frac{c^{(k)}_{\pid}}{\sqrt2} | \pit \rangle +
i\, \frac{c^{(k)}_{\rhodpar}}{\sqrt2} | \rhocpar
\rangle - i\, \frac{c^{(k)}_{\rhodpar}}{\sqrt2} |
\rhotpar \rangle\ , \  k=2,4\ .\ \
\label{eqna}
\end{eqnarray}
For definiteness, let us take $m_{\parallel}^{(1)}<m_{\parallel}^{(3)}$,
$m_{\parallel}^{(2)}<m_{\parallel}^{(4)}$. Since for $\alpha=0$ and $B=0$
the Lagrangian shows an approximate symmetry under ${\rm
SU(2)}_{A}\otimes{\rm U(1)}_{A}$ chiral transformations, spontaneous
symmetry breaking leads to four pseudo-Goldstone bosons, viz.\ the three
pions and the $\eta$ meson. In the presence of the magnetic field, chiral
symmetry is explicitly broken from ${\rm SU(2)}_{A}\otimes{\rm U(1)}_{A}$
down to ${\rm U(1)}_{T^{3},\,A}\otimes{\rm U(1)}_{A}$; thus, one still has
two neutral mesons ---combinations of the neutral pion and the $\eta$---
that remain as pseudo-Goldstone bosons. Moreover, according to the previous
discussion, the latter must be pure $u$ and $d$-states. Since they should be
approximate mass eigenstates, one expects to find
$(c_{\piu}^{(1)},c_{\rhoupar}^{(1)})\approx(c_{\pid}^{(2)},c_{\rhodpar}^{(2)})\approx(1,0)$
and
$(c_{\piu}^{(3)},c_{\rhoupar}^{(3)})\approx(c_{\pid}^{(4)},c_{\rhodpar}^{(4)})\approx(0,1)$.
On the other hand, for $\alpha\neq0$ the presence of the 't Hooft term
introduces flavor mixing at the level of scalar and pseudoscalar four-quark
interactions, breaking the ${\rm U(1)}_{A}$ symmetry. Thus, the spin-isospin
decomposition gets the more general form given in Eq.~(\ref{decompara}),
where the lightest state can still be identified as an approximate Goldstone
boson. When $\alpha$ approaches $0.5$, the $\tilde{\eta}$ mass goes to
infinity and, accordingly, the $\left|\pic\right\rangle$ component in
Eq.~(\ref{decompara}) disappears from the remaining states.

In Table~\ref{taba} we quote the composition of the mass eigenstates $\tilde
M$ described in Fig.~\ref{Fig3}, for some representative values of the
magnetic field. For completeness, the coefficients corresponding to both
spin-isospin and spin-flavor basis are included. We note that while the mass
eigenvalues do not depend on whether $B$ is positive or negative, the
corresponding eingenvectors do. The relative signs in Table~\ref{taba}
correspond to the choice $B>0$.
\begin{table}[h]
\begin{center}
\begin{tabular}{ccccccccccccc}
\hline \hline
      State         &$\ $ &  $eB\ [{\rm GeV}^2]$
  &$\ $&\multicolumn{4}{c}{Spin-isospin composition}& &\multicolumn{4}{c}{Spin-flavor composition}\\
  &     &               &    & \hspace{3mm} $c^{(k)}_{\pic}$\hspace{3mm} & \hspace{3mm}$c^{(k)}_{\pit}$\hspace{3mm} & \hspace{3mm}$c^{(k)}_{\rhocpar}$\hspace{3mm} &\hspace{3mm} $c^{(k)}_{\rhotpar}$ \hspace{3mm}
      &$\ $ & \hspace{3mm} $c^{(k)}_{\piu}$\hspace{3mm} & \hspace{3mm}$c^{(k)}_{\pid}$\hspace{3mm} & \hspace{3mm}$c^{(k)}_{\rhoupar}$\hspace{3mm} &\hspace{3mm} $c^{(k)}_{\rhodpar}$ \hspace{3mm}\\
\hline
 $\tilde\pi\ ( k=1 )$         & & 0.05 & &   0.0037           &    0.9998         &  -0.0203          &    -0.0068     && 0.7096 & -0.7043 & -0.0192 & -0.0095      \\
                           & & 0.5  & &   0.1019           &    0.9910         &  -0.0822          &    -0.0285     && 0.7728 & -0.6287 & -0.0783 & -0.038      \\
                           & & 1.0  & &   0.1566           &    0.9841         &  -0.0797          &    -0.0274     && 0.8066 & -0.5851 & -0.0757 & -0.037       \\
\hline
 $\tilde\eta\ ( k=2 )$        & & 0.05 & &   0.9899           &   -0.0413         &  -0.0381          &    -0.1301     && 0.6708 & 0.7292  & -0.1189 & 0.0651      \\
                           & & 0.5  & &   0.8661           &   -0.3246         &   0.0582          &    -0.3757     && 0.3829 & 0.8420  & -0.2245 & 0.3068      \\
                           & & 1.0  & &   0.8353           &   -0.3445         &   0.1048          &    -0.4154     && 0.3470 & 0.8342  & -0.2196 & 0.3678      \\
\hline
 $\tilde\omega\ ( k=3 )$      & & 0.05 & &  -0.1979           &    0.2693         &   0.7601          &    -0.5572     && 0.0505 & -0.3304 & 0.1435  & 0.9315     \\
\hline
 $\tilde\rho\ ( k=4)$         & & 0.05 & &   0.4925           &    0.3312         &   0.4685          &     0.6544     && 0.5824 & 0.1141  & 0.7940  & -0.1315     \\
\hline \hline
\end{tabular}
\caption{  Composition of the $S_{z}=0$ meson mass eigenstates for
some selected values of $eB$. Results correspond to $\alpha =
0.1$. Relative signs hold for the choice $B>0$.} \label{taba}
\end{center}
\end{table}

Let us first discuss the composition of the $\tilde\pi$ state ($k=1$), which
is the one that has the lowest mass. We see that even though $\alpha$ is
relatively small, the effect of flavor mixing is already very strong; the
spin-isospin composition is clearly dominated by the $\pit$ component, which
is given by an antisymmetric equal-weight combination of $u$ and $d$ quark
flavors. Thus, the mass states are far from satisfying the flavor
disentanglement expected for the case $\alpha=0$ [see Eqs.~(\ref{eqna})], in
which one has two approximate Goldstone bosons. In fact, once $\alpha$ is
turned on, explicitly breaking the U(1)$_A$ symmetry, $\pit$ is the only
state that remains being a pseudo-Goldstone boson; this forces the
lowest-mass state $\tilde\pi$ to be dominated by the $\pit$ component. As
discussed above, the presence of the magnetic field distinguishes between
flavor components $\piu$ and $\pid$ instead of isospin states. However, it
is found that even for values of $\alpha$ as small as 0.01 the mass state
$\tilde\pi$ is still dominated by the $\pit$ component ($|c^{(1)}_{\pit}|^2
\gtrsim 0.9$) for the full range of values of $eB$ considered here. In other
words, extremely large magnetic fields would be required in order to rule
the composition of light mass eigenstates, which is otherwise dictated by
the invariance under ${\rm U(1)}_{T^{3},\,A}$ transformations. Coming
back to the case $\alpha = 0.1$, we see that, although relatively small, the
effect of the magnetic field on the composition of the $\tilde\pi$ state can
be observed from the values in Table~\ref{taba}. When $eB$ gets increased,
it is found that there is a slight decrease of the component $\pi^3$ in
favor of the others. In addition, a larger weight is gained by the
$u$-flavor components, as one can see by looking at the entries
corresponding to the spin-flavor states (last four columns of
Table~\ref{taba}): one has $|c^{(1)}_{\piu}|^2+|c^{(1)}_{\rhoupar}|^2 = 0.50
(0.66)$ for $eB=0.05 (1.0)$~GeV$^2$. This can be understood  noticing that
the magnetic field is known to reduce the mass of the lowest neutral meson
state~\cite{Luschevskaya:2014lga,Bali:2017ian,Ding:2020hxw}. Thus, for large
$eB$ it is expected that $\tilde\pi$ will have a larger component of the
quark flavor that couples strongly to the magnetic field (i.e., the $u$
quark). Concerning the vector meson components of the $\tilde\pi$ state, it
is seen that they are completely negligible at low values of $eB$, reaching
a contribution $|c^{(1)}_{\rhoupar}|^2+|c^{(1)}_{\rhodpar}|^2 \simeq 0.01$
(i.e., about $1\%$) at $eB= 1$~GeV$^2$.

Turning now to the composition of the $\tilde\eta$ state ($k=2$) in
Table~\ref{taba}, we see that, as expected from the above discussion, it is
dominated by the $\pi^0$ ($I=0$) component for values of $eB$ up to
1~GeV$^2$. Regarding the flavor composition, in this case the $d$-quark
content is the one that increases as $eB$ does, with
$|c^{(2)}_{\pid}|^2+|c^{(2)}_{\rhodpar}|^2 = 0.54 (0.70)$ for $eB=0.05
(1.0)$~GeV$^2$. Now the weight of the vector components is larger than in
the case of the $\tilde\pi$ state,
$|c^{(1)}_{\rhoupar}|^2+|c^{(1)}_{\rhodpar}|^2$ ranging from $0.02$ at $eB=
0.05$~GeV$^2$ to $0.17$ at $eB= 1.0$~GeV$^2$. This is probably due to the
fact that for $\alpha = 0.1$ the $\tilde\eta$ mass is closer to vector meson
masses.

Finally, let us comment on the composition of the $\tilde\omega$ and
$\tilde\rho$ states ($k=3$ and $k=4$, respectively). As mentioned above, the
masses of these states reach the threshold for $q\bar q$ decay for rather
low values of the magnetic field, hence our predictions for these quantities
should be taken as qualitative ones for a major part of the $eB$ range
considered here. It is worth noticing that there is a multiple number of
thresholds, which get successively opened each time the meson mass is
sufficiently large so that the quark and antiquark meson components can
populate a new Landau level. The first thresholds in the $\bar{u}u$ and the
$\bar{d}d$ sectors are reached at meson masses equal to $2M_{u}$ and $2M_d$,
respectively. It is important to realize that they do not correspond to a
free quark together with a free antiquark, but to the quark and antiquark in
their lowest Landau levels. Taking $B>0$, if both the quark and the
antiquark have vanishing $z$ component of the momentum, the corresponding
spin configurations are
$u\left(S_{z}=+\frac{1}{2}\right)\,\bar{u}\left(S_{z}=-\frac{1}{2}\right)$
and
$d\left(S_{z}=-\frac{1}{2}\right)\,\bar{d}\left(S_{z}=+\frac{1}{2}\right)$.
In both cases, the magnetic dipole moments of the quark and the antiquark
are parallel to the magnetic field; the difference between both
configurations arises from the opposite signs of the quark electric charges.
We only quote in Table~\ref{taba} the $\tilde\omega$ and $\tilde\rho$
compositions in the presence of a low magnetic field $eB=0.05$~GeV$^2$, for
which the masses of both states are below the $2M_f$ threshold and the
values of the coefficients $c^{(k)}_M$ should be more reliable.
Interestingly, we note that even at this low value of the magnetic field the
composition of the vector meson mass states is clearly flavor-dominated:
from Table~\ref{taba} one has $|c^{(3)}_{\pid}|^2+|c^{(3)}_{\rhodpar}|^2 =
0.98$, $|c^{(4)}_{\piu}|^2+|c^{(4)}_{\rhoupar}|^2 = 0.97$. Thus, whereas for
no external field one usually identifies the (approximately degenerate) mass
states as isospin eigenstates $\rho^0$ and $\omega$, in the presence of the
magnetic field the states $\tilde\rho$ and $\tilde\omega$ are closer to a
$\rhoupar$ and a $\rhodpar$, rather than a $\rhotpar$ and a $\rhocpar$. In
fact, given the symmetry of the vector-like interactions in the Lagrangian
in Eq.~(\ref{lagrangian}), the small deviation of $\tilde\rho$ and
$\tilde\omega$ from pure flavor states can be attributed to the mixing with
the pseudoscalar sector, where isospin states are dominant. Notice that
although the vector components are larger than the pseudoscalar ones, the
weight of the latter is not negligible, specially for the $\tilde\rho$ state
(which is the one with a larger mass, as shown in Fig.~\ref{Fig3}), with
$|c^{(4)}_{\piu}|^2+|c^{(4)}_{\pid}|^2\simeq 0.35$. This can be understood
from an analysis similar to the one performed for the meson mass thresholds
in terms of the quark spins. A larger content of the
$d\left(S_{z}=-\frac{1}{2}\right)\,\bar{d}\left(S_{z}=+\frac{1}{2}\right)$
component has to be expected in the case of the $\tilde{\omega}$, while
there should be a larger content of the
$u\left(S_{z}=+\frac{1}{2}\right)\,\bar{u}\left(S_{z}=-\frac{1}{2}\right)$
one in the case of the $\tilde{\rho}$. From Table I it is seen that these
combinations correspond to $\big(c_{\pid}^{(3)} -
c_{\rhodpar}^{(3)}\big)/\sqrt{2}=-0.89$ for the $\tilde{\omega}$ and $\big(
c_{\piu}^{(4)}+c_{\rhoupar}^{(4)} \big)/\sqrt{2}=0.97$ for the
$\tilde{\rho}$, under a magnetic field as low as $e B=0.05$~GeV$^{2}$ ---and
this effect should be more significant for larger values of $eB$.

\begin{figure}[h]
     \centering{}\includegraphics[width=1.\textwidth]{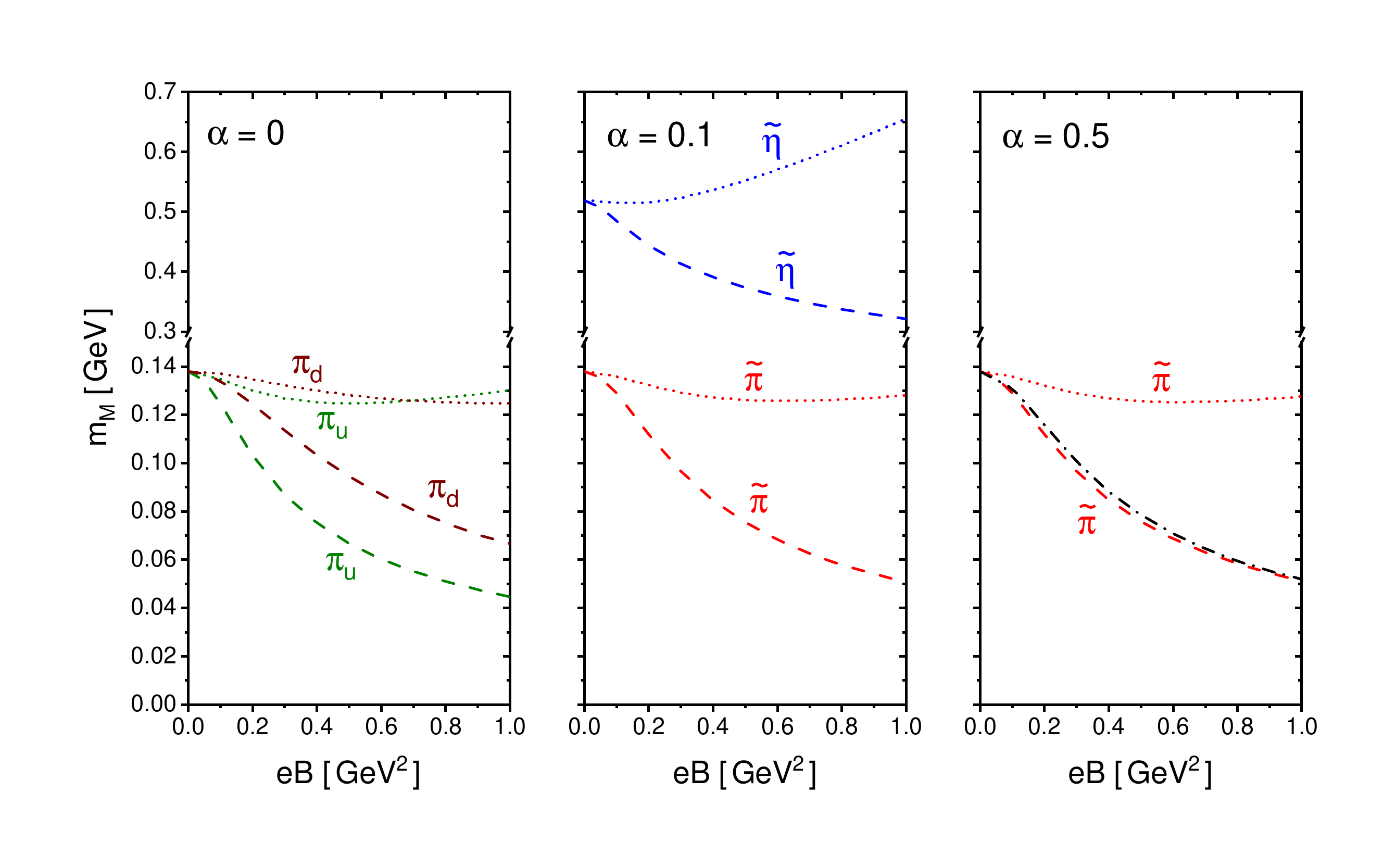}
\caption{(Color online) Masses of the lightest $S_z =0$ mesons as
functions of $eB$, for various values of $\alpha$. Dashed (dotted) lines
correspond to the case in which the mixing between pseudoscalar and vector
states is (is not) included. The dash-dotted line in the right panel is
obtained from the approximate expression in Eq.~(\ref{eqappc}).}
\label{Fig4}
\end{figure}

We analyze in what follows the impact of both flavor mixing and
pseudoscalar-vector mixing on the masses of the  lightest states. In fact,
this is one of the main issues of this work. In Fig.~\ref{Fig4} we show the
$B$ dependence of light meson masses with (dashed lines) and without (dotted
lines) pseudoscalar-vector mixing, considering three representative values
of the flavor-mixing constant $\alpha$.  The results without
pseudoscalar-vector mixing are obtained just by setting to zero the
off-diagonal polarization functions $\hat
{\jnomatrix}_{\pia\rhobmu}^{\rm mag}$ and $\hat
{\jnomatrix}_{\rhoamu\pib}^{\rm mag}$ in Eq.~(\ref{jregjmag}). Let us
focus on $m_{\tilde\pi}$, considering first the effect of varying $\alpha$;
as can be seen from Fig.~\ref{Fig4}, this effect is rather independent of
whether pseudoscalar states mix with vectors or not.
We observe that for $\alpha=0$ (no flavor mixing) there are two light mesons
having similar masses; as stated above, these are pure flavor states and can
be identified as approximate Goldstone bosons. For $\alpha\neq 0$, the mass
of the $\tilde{\pi}$ state is still protected owing to its pseudo-Goldstone
boson character, whereas the $\tilde{\eta}$ state becomes heavier when
$\alpha$ gets increased, and disappears from the spectrum in the limit
$\alpha=0.5$.

{}From Fig.~\ref{Fig4} it is also seen that, for all values of $\alpha$, the
mixing between pseudoscalar and vector meson states produces a significant
decrease in the mass of the lightest state. This might be surprising, since
---as shown above--- the vector meson components of the $\tilde{\pi}$
state are found to be very small even for large values of $eB$. The
explanation of this puzzle is discussed in detail in App.~C, where it is
shown that these two facts are indeed consistent. Moreover, for $\alpha=0.5$
it is shown that if the pseudoscalar-vector meson mixing is treated
perturbatively, one can derive a simple formula for the $B$ dependence of
the $\tilde\pi$ mass, viz.
\begin{equation}
m_{\tilde\pi} \ = \ \frac{\bar m_{\tilde\pi}}{\sqrt{1 +
\kappa \, (\bar m_{\tilde\pi}\,
eB)^2/M_f}}\ ,
\label{eqappc}
\end{equation}
where $\kappa = 5 N_c^2\, g\, g_{v}/(18 \pi^4 m_c)$, $f$ is either $u$ or
$d$, and $\bar m_{\tilde\pi}$ stands for the $\tilde\pi$ mass when no mixing
is considered. Taking into account that $\bar m_{\tilde\pi}$ is very weakly
dependent on $B$ (see dotted lines in Fig.~\ref{Fig4}), it follows that
$m_{\tilde\pi}$ basically depends on the magnetic field through the ratio
$(eB)^2/M_f$. Notice that the $B$ dependence of $M_f$ for $\alpha=0.5$ is
represented by the solid line in Fig.~\ref{Fig1}. The numerical results for
$m_{\tilde\pi}$ from Eq.~(\ref{eqappc}), within the approximation $\bar
m_{\tilde\pi} = m_\pi(B=0)$ [see Eq.~(\ref{eqappb})] are indicated by the
black dash-dotted line in the right panel (corresponding to $\alpha=0.5$) of
Fig.~\ref{Fig4}. It can be seen that they are in excellent agreement with
those obtained from the full calculation.

\begin{figure}[h]
    \centering{}\includegraphics[width=0.75\textwidth]{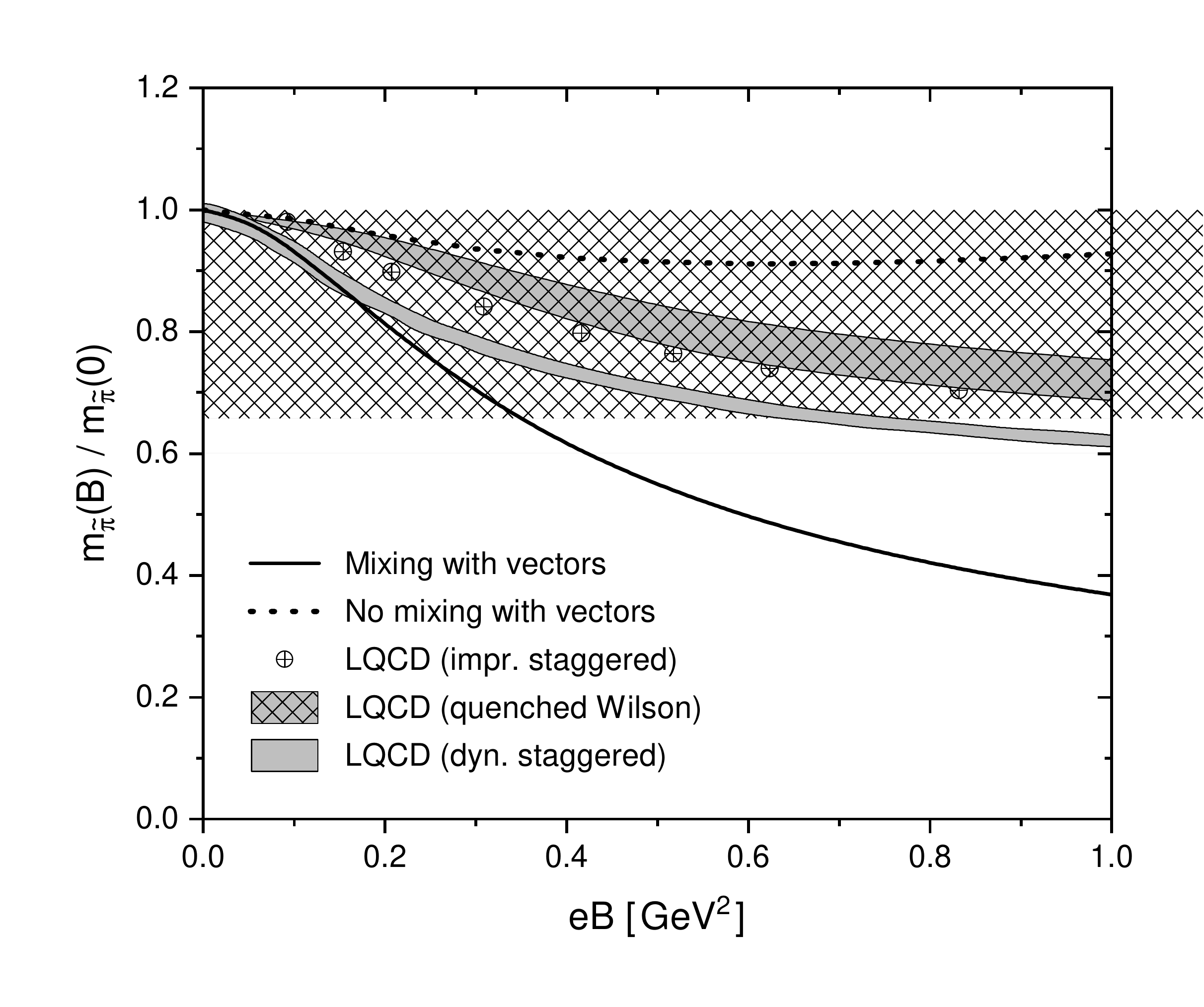}
\caption{Normalized mass of the $\tilde\pi$ meson (lightest state of the
$S_z=0$ sector) as a function of $eB$, compared with LQCD results quoted in
Ref.~\cite{Bali:2017ian}(quenched Wilson fermions), Ref.~\cite{Ding:2020hxw}
(improved staggered quarks) and
Refs.~\cite{Bali:2017ian,Borsanyi:2010cj,Bali:2011qj} (dynamical staggered
quarks). Solid and dotted lines correspond to NJL results with and without
pseudoscalar-vector meson mixing, respectively.}
\label{Fig5}
\end{figure}

To conclude this subsection, in Fig.~\ref{Fig5} we compare our results for
the mass of the $\tilde\pi$ state with those obtained in LQCD calculations,
reported in Ref.~\cite{Bali:2017ian} (quenched Wilson fermions),
Ref.~\cite{Ding:2020hxw} (improved staggered quarks) and
Refs.~\cite{Bali:2017ian,Borsanyi:2010cj,Bali:2011qj} (dynamical staggered
quarks). We first note that in those calculations the authors neglect
disconnected diagrams as well as the associated mixing, and work with the
individual flavor states instead. In our calculation this can be achieved by
setting $\alpha=0$. In any case, as seen from the above analysis, the mass
of the lightest meson is approximately independent of the value of $\alpha$;
therefore, it is reasonable to compare the mentioned LQCD results with those
obtained using the reference value $\alpha=0.1$ that leads to an acceptable
value for the $\eta$ meson mass at vanishing external field. We also note
that LQCD results have been obtained using different methods and values of
the pion mass at $B=0$. In particular, the most recent ones (i.e.~those in
Ref.~\cite{Ding:2020hxw}) are based on a highly improved staggered quark
action that uses $m_\pi(B=0) = 220$~MeV, while the calculations in
Refs.~\cite{Bali:2017ian,Borsanyi:2010cj,Bali:2011qj} take the physical
value of $m_\pi$ within a staggered simulation setup. Anyway, in our model
we see that when the pseudoscalar-vector meson mixing is included, the
values for the $\tilde\pi$ meson mass lie in general below LQCD predictions.
We have checked that this general result is quite insensitive to a
reasonable variation of the model parameters. In addition, we have verified
that the situation does not change significantly if the $B=0$ expressions
are regularized using the Pauli-Villars scheme, as proposed e.g.\ in
Ref.~\cite{Avancini:2022qcp}.

\subsection{$S_z=\pm 1$ vector meson sector}

In this subsection we present the numerical results associated with the
coupled system composed by the neutral vector mesons with $|S_{z}|=1$. As
discussed in Sect.~\ref{neutral}, for any value of $\alpha$ the mass
eigenstates can be identified according to their flavor content,
$|\rhouperp\rangle$ and $|\rhodperp\rangle$. The corresponding
masses can be obtained by solving the equations
$G_{f\perp}(-m_{\rho_{f\perp}}^{2})=0$, for $f=u,d$, with
$G_{f\perp}(-m^{2})$ given by Eq.~(\ref{gfperp}).

\begin{figure}[h]
    \centering{}\includegraphics[width=0.75\textwidth]{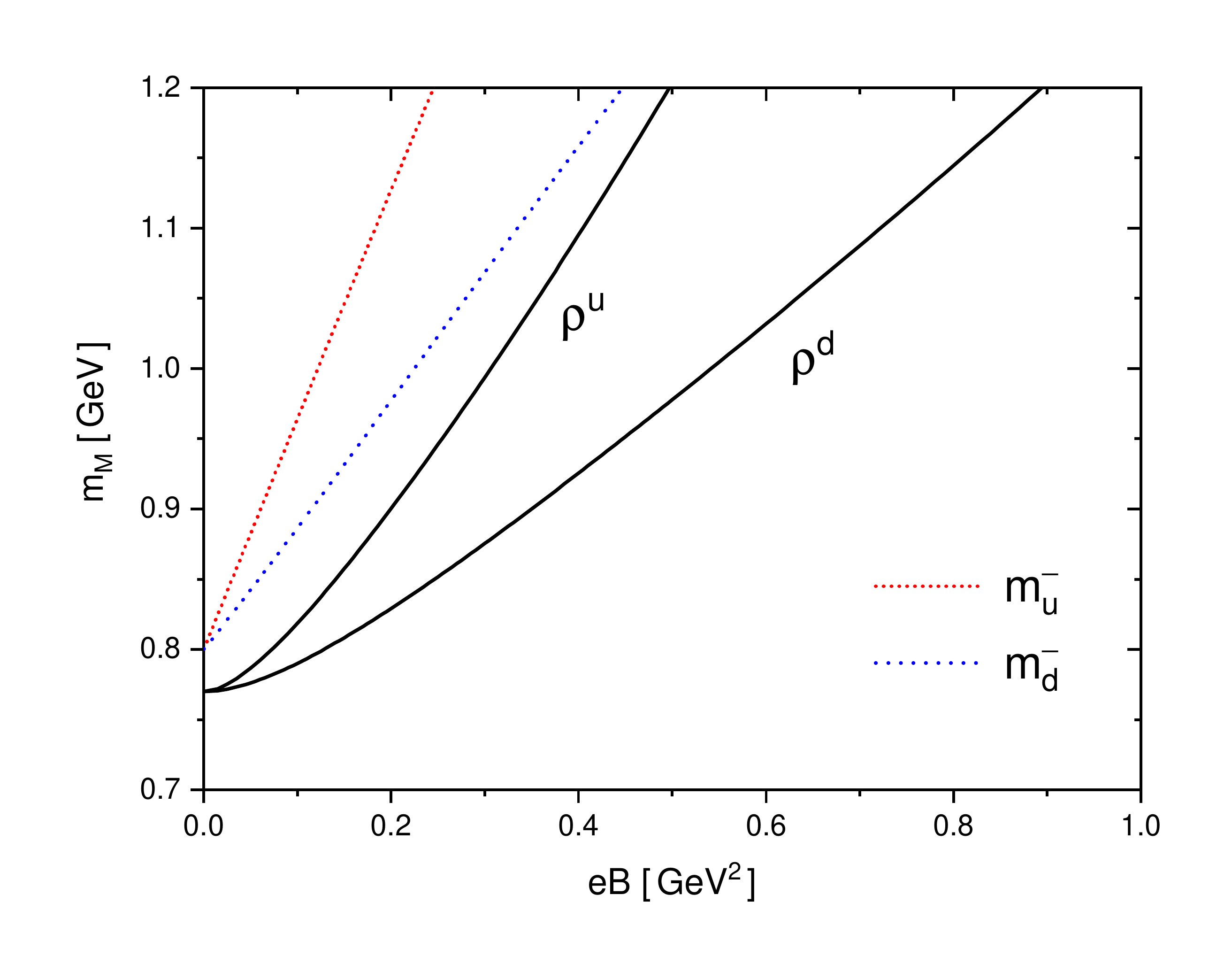}
\caption{(Color online) Masses of the $S_z = \pm 1$ vector meson states as
functions of $eB$. Dotted and short-dotted lines indicate $m_d^-$ and
$m_u^-$ quark-antiquark production thresholds, respectively.}
\label{Fig6}
\end{figure}

The numerical results for the meson masses as functions of the
magnetic field for the case $\alpha=0.1$ are shown in
Fig.~\ref{Fig6}, where it is seen that both $m_{\rhouperp}$
and $m_{\rhodperp}$ get increased with $B$. The enhancement
is larger in the case of the $\rhouperp$ mass; this can be
understood from the larger (absolute) value of the $u$-quark
charge, which measures the coupling with the magnetic field. As
in the case of $S_z = 0$ mesons, there are multiple mass
thresholds for $q\bar q$ pair production [see Eqs.~(\ref{i4f}) and
(\ref{emes})]. The lowest one, reached at $m_{\rho_{f\perp}} =
m_f^- = M_{f}+\sqrt{M_{f}^{2}+2B_{f}}$, corresponds now to the
situation in which {\it both} the spins of the quark and the
antiquark components of the $\rho_{f\perp}$ are aligned (or
anti-aligned) with the magnetic field. Notice that in this case
one of the fermions lies in its lowest Landau level, while the
other one is in the first excited Landau level; whether both
particle spins are aligned or anti-aligned with the magnetic field
depends on the signs of $S_{z}$ and $B$. It can be seen that the
values of $m_f^-$ for $f=u$ or $d$ are not surpassed by the
corresponding meson masses $m_{\rho_{f\perp}}$ in the studied
region, and consequently these masses are found to be smooth real
functions of $eB$, as shown in Fig.~\ref{Fig6}. We stress that the
mass values $m=2M_u$ and $m=2M_d$ are not actual thresholds in
this case, since ---as discussed above--- they correspond to
lowest Landau level quark configurations that lead to $S_z=0$
meson states. The absence of these thresholds can be formally
shown by looking at the expression in Eq.~(\ref{gfperp}); it can
be seen that although the functions $I_{2f}(-m^{2})$ and
$I_{4f}^{{\rm mag}}(-m^{2})$ become complex for $m>2M_{f}$,
imaginary parts cancel each other and one ends up with a vanishing
absorptive contribution.

\begin{figure}[h]
   \centering{}\includegraphics[width=1\textwidth]{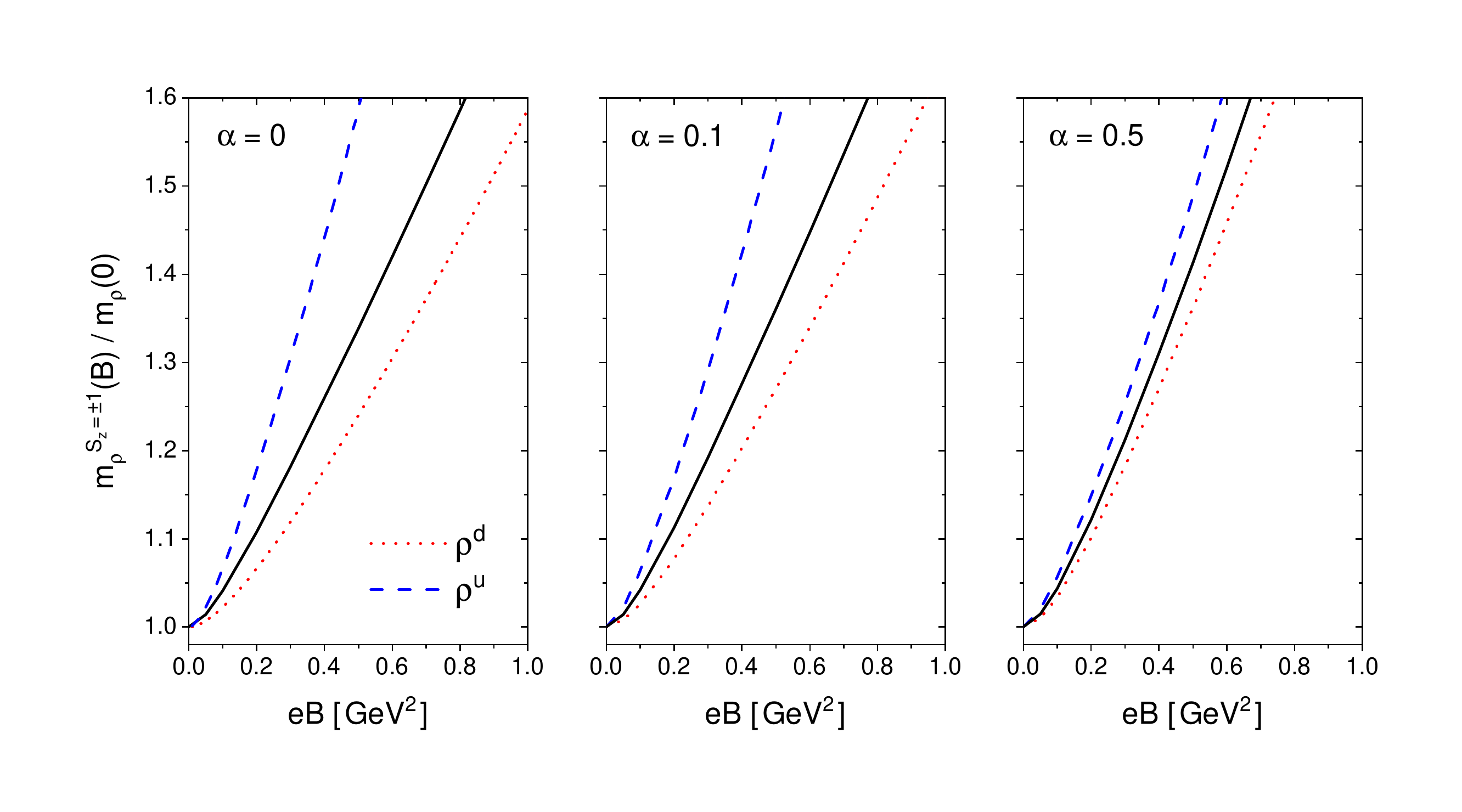}
\caption{(Color online) Masses of $S_z=\pm 1$ vector mesons as functions of
$eB$, for various values of $\alpha$. The full black lines correspond to the
case in which the mixing between pseudoscalar and vector meson states is not
considered.} \label{Fig7}
\end{figure}

It should be pointed out that even though there is no direct flavor mixing
in this sector, $\rhouperp$ and $\rhodperp$ meson masses still depend
on $\alpha$. This is due to the fact that the values of $M_u$ and $M_d$
obtained at the MF level get modified by flavor mixing. We recall that
$M_u(eB) = M_d(2eB)$ for $\alpha=0$, while for $\alpha=0.5$ one has $M_u(eB)
= M_d(eB)$. The effect of flavor mixing is illustrated in Fig.~\ref{Fig7},
where we show the $B$ dependence of $\rhouperp$ and $\rhodperp$ meson
masses for $\alpha=0$, 0.1 and 0.5. As expected from the aforementioned
relations between $M_u$ and $M_d$, it is seen that the curves for both
masses tend to become more similar as $\alpha$ increases. However, the
overall effect is found to be relatively weak. As a reference we also plot
(full black line) the situation in which the mixing between $S_z=\pm 1$
vector states is neglected, and, therefore, the masses of both states
coincide. We see that even in the case $\alpha=0.5$ there is a certain
non-negligible repulsion between states when the mixing term is turned on.

\begin{figure}[h]

\centering{}\includegraphics[width=0.75\textwidth]{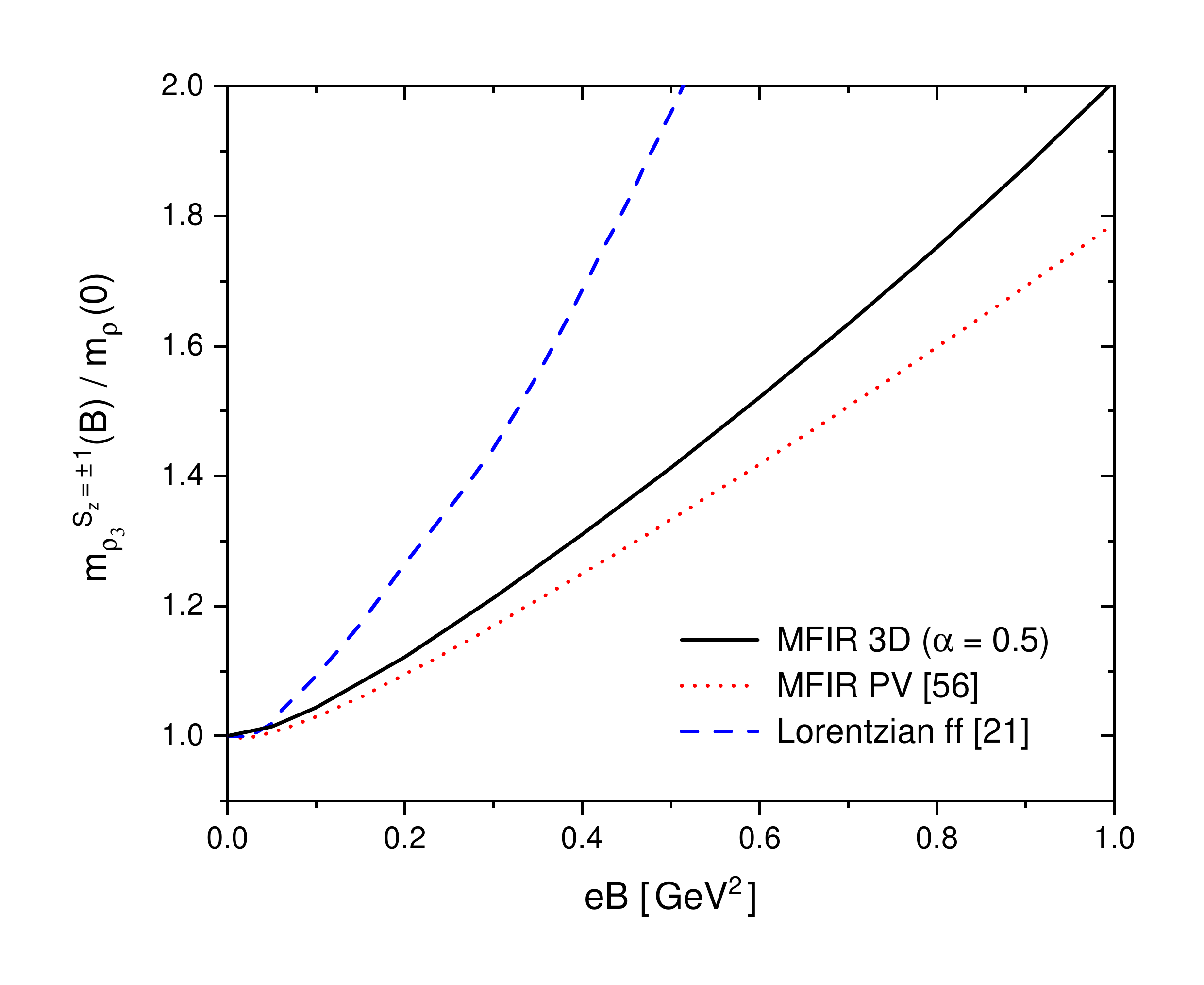}
\caption{(Color online) Mass of the $\rho$ meson with $S_z=\pm 1$ for the
case in which and $\alpha = 0.5$ and there is no mixing between pseudoscalar
and vector meson states. Results quoted in the literature using other
regularization methods are also shown.}
\label{Fig8}
\end{figure}

It is also interesting at this stage to analyze the impact of the
regularization procedure on the predictions of the model. In Fig.~\ref{Fig8}
we show our results for the $\rhotperp$ mass together with those obtained in
Ref.~\cite{Liu:2014uwa} and Ref.~\cite{Avancini:2022qcp}. To carry out a
proper comparison, in our model we have taken $\alpha=0.5$ and have set to
zero the $\rhocperp-\rhotperp$ mixing contributions, as done in those works
(in which the $\rhocperp$ state is not included). Notice that this case
corresponds to the solid line in the right panel of Fig.~\ref{Fig7}. In
Ref.~\cite{Liu:2014uwa}, divergent integrals are regularized through the
introduction of Lorenztian-like form factors, both for vacuum and
$B$-dependent contributions. On the other hand, in
Ref.~\cite{Avancini:2022qcp} the regularization is carried out using the
MFIR method, as in the present work. However, to deal with vacuum-like terms
the authors of Ref.~\cite{Avancini:2022qcp} choose a Pauli-Villars
regularization, instead of the 3D-cutoff scheme considered here. From
Fig.~\ref{Fig8} it is seen that our results for $m_{\rhotperp}$ (black solid
line) are quite similar to those found in Ref.~\cite{Avancini:2022qcp} (red
dotted line), indicating that they are not too much sensitive to the
prescription used for the regularization of vacuum-like terms, once the MFIR
method is implemented. Meanwhile, the $\rhotperp$ mass obtained by means of
a form factor regularization (blue dashed line) shows a much stronger
dependence on the magnetic field, specially for large values of $eB$. These
results are consistent with those found in Ref.~\cite{Avancini:2019wed} for
the regularization scheme dependence of the condensates in the presence of
the magnetic field.

\begin{figure}[h]
     \centering{}\includegraphics[width=0.75\textwidth]{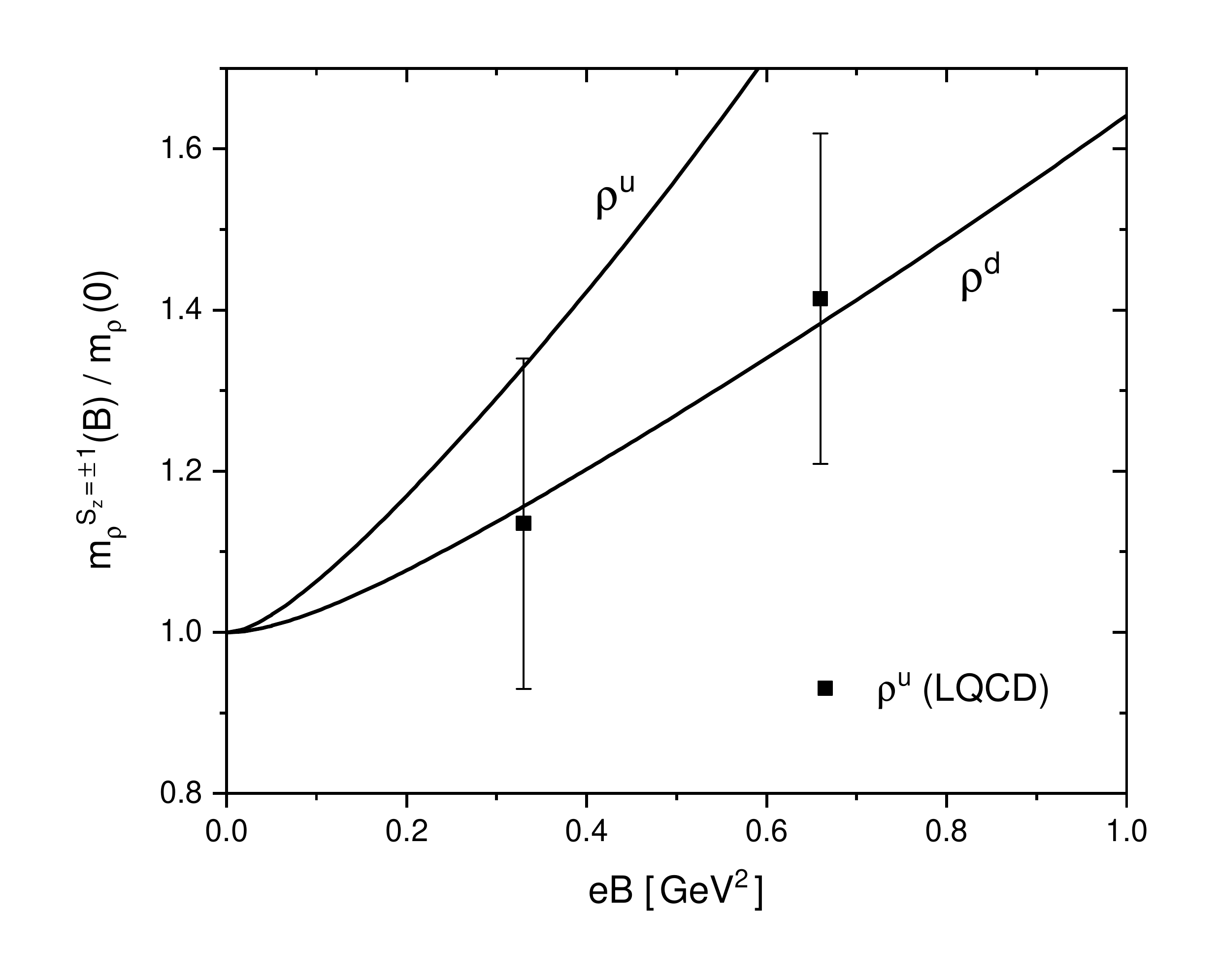}
\caption{(Color online) Masses of $S_z=\pm 1$ vector meson states for $\alpha =
0.1$, compared with LQCD results given in Ref.~\cite{Bali:2017ian}.}
\label{Fig9}
\end{figure}

Finally, in Fig.~\ref{Fig9} we compare our results for the case $\alpha=0.1$
(dashed and dotted lines in the central panel of Fig.~\ref{Fig7})
with those quoted in Ref.~\cite{Bali:2017ian} for the $\rhouperp$ mass
using LQCD calculations. In fact, these lattice results are obtained
for a large vacuum pion mass of about $400~$MeV; the comparison still makes
sense, however, since we have checked that our results are rather robust
under changes in the current quark masses leading to such a large value of
$m_\pi$. Considering the large error bars, from the figure one observes that
LQCD results seem to indicate an enhancement of $m_{\rhouperp}$ when the
magnetic field is increased, in agreement with the predictions from the NJL
model. This qualitative behavior has been also found in previous LQCD
studies~\cite{Luschevskaya:2012xd,Luschevskaya:2014lga,
Luschevskaya:2015bea,Andreichikov:2016ayj}.

\subsection{$B$-dependent four-fermion couplings}
\label{bdependent}

As mentioned in the Introduction, while local NJL-like models are able to
reproduce the magnetic catalysis (MC) effect at vanishing temperature, they
fail to describe the so-called inverse magnetic catalysis (IMC) observed in
lattice QCD. Among the possible ways to deal with this problem, one of the
simplest approaches is to allow the model coupling constants to depend on
the magnetic field. With this motivation, we explore in this subsection the
possibility of considering a magnetic field dependent coupling $g(eB)$. For
definiteness, we adopt for this function the form proposed in
Ref.~\cite{Avancini:2016fgq}, namely
\begin{equation}
g(eB) \ = \ g\, {\cal F}(eB)\ ,
\label{gdeb}
\end{equation}
where
\begin{equation}
{\cal F}(eB) =  \kappa_1 + (1-\kappa_1) \, e^{-\kappa_2\, (eB)^2}\ ,
\label{fdeb}
\end{equation}
with $\kappa_1=0.321$, $\kappa_2=1.31$~GeV$^{-2}$. Assuming this form for
$g(eB)$, the effective quark masses are found to be less affected by the
presence of the magnetic field than in the case of a constant $g$. In fact,
they show a non-monotonous behavior for increasing $B$, resembling the
results found in Refs.~\cite{Endrodi:2019whh,Avancini:2021pmi}. It should be
stressed that in spite of the rather different behavior of the dynamical
quark masses, a similar zero-temperature magnetic catalysis effect is
obtained both for a constant $g$ and for a variation with $B$ of the
form given by Eq.~(\ref{fdeb}).

\begin{figure}[h]
     \centering{}\includegraphics[width=0.9\textwidth]{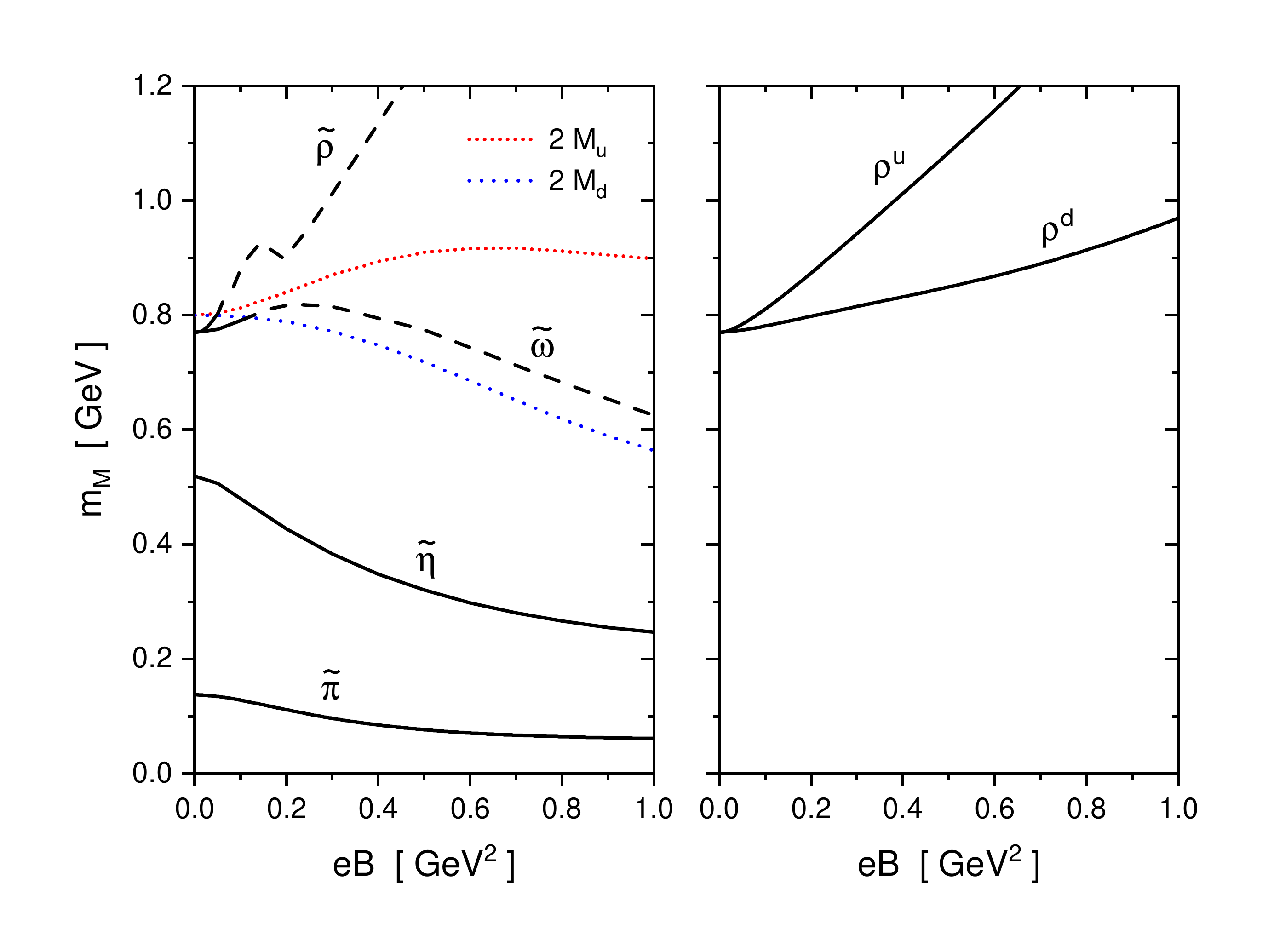}
\caption{(Color online) Left (right) panel: Masses of the $S_z=0$ ($S_z=\pm
1)$ meson states as functions of $eB$, for $B$-dependent couplings $g(eB)/g=
g_v(eB)/g_v = {\cal F}(eB)$ [see Eq.~(\ref{fdeb})]. The results correspond
to the case $\alpha = 0.1$.}
\label{Fig10}
\end{figure}

Regarding the vector meson sector, one has to choose some assumption for the
$B$ dependence of the vector coupling constant. One possibility is to
suppose that, due to their common gluonic origin, the vector couplings are
affected by the magnetic field in the same way as the scalar and
pseudoscalar ones. That is to say, one could take $g_{v}(eB) = g_{v}\, {\cal
F}_v(eB)$, with ${\cal F}_v(eB) = {\cal F}(eB)$. Under these assumptions, we
have obtained numerical results for the behavior of meson masses with the
magnetic field. The curves for the case $\alpha=0.1$ are given in
Fig.~\ref{Fig10}, where we also show the $q\bar q$ production
thresholds (dotted lines).

By comparison with the results in Fig.~\ref{Fig3} and Fig.~\ref{Fig6}, it
can be observed that the $B$ dependence of the couplings has a significant
qualitative effect only in the case of the $\tilde\omega$ state. It is found
that the mass of this state follows quite closely the position of the lowest
$q\bar q$ production threshold, $2 M_d$, which ---as stated--- does get
affected by the $B$ dependence of $g$. The behavior of the masses of the
other mesons do not change qualitatively with respect to the case $g={\rm
constant}$, and something similar happens with their composition and their
dependence on $\alpha$. In particular, the results for the ratio
$r_\pi={m_{\tilde \pi}}(eB)/m_\pi(0)$ are almost identical to those obtained
in Sect.~\ref{sec_sz0} (solid line in Fig.~\ref{Fig5}).

Given the fact that $g_v(eB)$ is not so well constrained as in the case of
the scalar coupling, one can, in principle, introduce a new function ${\cal
F}_v(eB)$, different from ${\cal F}(eB)$. The freedom in the election of
this function can be used to reproduce the results for the ratio $r_\pi$
obtained through LQCD calculations. It can be seen, however, that in this
case the masses of the $S_z=\pm 1$ vector mesons increase even faster than
in the case in which $B$-independent couplings are used.

\section{Conclusions}

In this work we have studied the mass spectrum of light neutral pseudoscalar
and vector mesons in the presence of an external uniform magnetic field
$\vec{B}$. For this purpose we have considered a two-flavor NJL-like model
in the Landau gauge. This model includes isoscalar and isovector couplings
in the scalar-pseudoscalar sector and in the vector sector. A flavor mixing
term in the scalar-pseudoscalar sector, regulated by a constant $\alpha$,
has also been included. For $\alpha=0$ there is not flavor mixing, but
flavor degeneracy gets broken by the magnetic field and $M_{u}\neq M_{d}$ ,
while for $\alpha=0.5$ one has maximum flavor mixing, as in the case of the
standard version of the NJL model, and in this case $M_{u}=M_{d}$. To
account for the usual divergences of the NJL model, we have considered here
the magnetic field independent regularization (MFIR) method, which has been
shown to reduce the dependence of the results on the model parameters. It
should be stressed that for neutral mesons the contributions to the
polarization functions arising from Schwinger phases in quark propagators
get cancelled; as a consequence, the polarization functions turn out to be
diagonal in the usual momentum basis.

It is important to note that the presence of an electromagnetic field allows
for isospin mixing. In addition, the axial character of the magnetic field
together with the loss of rotational invariance lead to pseudoscalar-vector
mixing. These mixing contributions are usually forbidden by isospin and
angular momentum conservation. However, they arise and may become important
in the presence of the external magnetic field. Although full rotational
invariance is broken, invariance under rotations around the magnetic field
direction survives. Therefore, the projection of the vector meson spin in
the field direction, $S_{z}$, is the observable that organizes the obtained
results. Our analysis shows that for the determination of the masses (i.e.,
if particles are taken at rest), the scalar mesons, which in our case
include the $f_{0}$ (or $\sigma)$ and $a_{0}^{0}$ states, mix with each
other but decouple from other mesons. Thus, they can be disregarded in the
analysis of the pseudoscalar and vector meson masses. The remaining meson
space can be separated into three subspaces: pseudoscalar and vector mesons
with $S_{z}=0,$ including $\pi^{0}$, $\eta$, $\rho^{0}$ and $\omega$, which
mix with each other; vector mesons with $S_{z}=+1,$ including $\rho^{0}$ and
$\omega$ mesons; same as before, with $S_{z}=-1.$

Regarding the $S_{z}=0$ sector, we observe two different behaviors for the
meson masses. The masses of the two lightest mesons, which we have called
$\tilde{\pi}$ and $\tilde{\eta}$, are determined by the underlying
symmetries and their breaking pattern. In the presence of the magnetic
field, with $\alpha=0,$ one has a ``residual'' ${\rm
U(1)}_{T^{3}}\otimes{\rm U(1)}_{T^{3},A}\otimes{\rm U(1)}_{A}$ chiral
symmetry, explicitly broken only by a (small) current mass term,
$m_{c}\neq0,$ which guarantees the pseudo-Goldstone character of these two
states. We have shown that flavor degeneracy gets broken by the magnetic
field and mass eigenstates are separated into particles with pure $u$ or $d$
quark content. For $\alpha=0.1,$ which leads to a reasonable value for the
$\eta$ mass in the absence of the magnetic field, the ${\rm U(1)}_{A}$
symmetry is broken and only one pseudo-Goldstone boson, $\tilde{\pi},$
survives. From our results, we can conclude that even for magnetic field
values as large as $eB=1\,\text{GeV}^{2},$ the $\tilde{\pi}$ state is mostly
a pseudoscalar isovector (third component) and $\tilde{\eta}$ is mostly
a pseudoscalar isoscalar. Increasing the magnetic field intensity from a low
value of $eB=0.05$~GeV$^{2}$ to $eB=1$~GeV$^{2}$ we observe that the $u$
content of the $\tilde{\pi}$ and the $d$ content of the $\tilde{\eta}$ get
enhanced.

On the other hand, regarding the quark structure of the two heaviest mesons,
which we call $\tilde{\omega}$ and $\tilde{\rho}$, it is found that even for
a low value of the magnetic field, $eB=0.05\,\text{GeV}^{2},$ the mass
eigenstates turn out to be clearly dominated by the quark flavor content and
spin orientation. This is what we could expect, since the magnetic field
tends to separate quarks according to their electric charges, and favors
that their magnetic moments be orientated parallel to the field direction.

The lack of confinement in the NJL model implies that the polarization
functions get absorptive contributions, related with $q\bar{q}$ pair
production, beyond certain thresholds. In the presence of the magnetic
field, the position of each threshold is flavor and spin dependent, in such
a way that for $S_{z}=0$ we have thresholds for meson mass values
$m_{f}=2\,M_{f}$, while for $S_{z}=\pm1$ the thresholds rise to higher
values $m_{f}=M_{f}+\sqrt{M_{f}^{2}+2\,B_{f}}.$ As a consequence, we find
that $\tilde{\omega}$ and $\tilde{\rho}$ states with $S_{z}=0$ enter into
the continuum for values of the magnetic field around $eB\sim0.1$~GeV$^{2},$
whereas $S_{z}=\pm1$ meson masses always lie under $q\bar q$ production
thresholds for the considered range of values of $eB$. A common result for
all these states is that their masses show an appreciable growth when the
magnetic field varies from zero to $eB=1$~GeV$^{2}.$ In the case of
$S_{z}=\pm1,$ the model reproduces reasonably well present LQCD results for
$\rho^{u}_\perp$, taking into account the uncertainties in LQCD simulations.

We have observed that the mass of the lightest state, $\tilde{\pi},$ gets
reduced as the magnetic field increases. This behavior reproduces the trend
of existing LQCD results. However, our results overestimate the mass
reduction as compared to the one found in LQCD simulations. It is seen that
this reduction is significantly affected by the mixing between pseudoscalar
and vector components, a fact that turns out to be independent of the value
of the flavor mixing parameter $\alpha$. From an analytical perturbative
analysis, we have carefully studied how a small value of the vector
components in the $\tilde{\pi}$ state can lead to a significant reduction of
its mass. It is seen that both the mixture of the $\pi$ channel with the
$\omega$ and $\rho$ channels contribute to this mass shrinkage.

While local NJL-like models are able to reproduce the magnetic catalysis
effect at vanishing temperature, they fail to lead to the so-called inverse
magnetic catalysis. One of the simplest ways to deal with this problem is to
allow that the model coupling constants depend on the magnetic field. With
this motivation, we have explored the possibility of considering magnetic
field dependent couplings $g(eB)$ and $g_v(eB)$. For definiteness we take
the same dependence on $B$ for both couplings; in that case, our results
show that, for any value of $\alpha$, the mass of the $\tilde{\omega}$ state
with $S_{z}=0$ is the only one that becomes significantly modified with
respect to the case in which $g$ and $g_v$ do not depend on the magnetic
field. In particular, the $B$-dependence of the ratio
$r_{\pi}=m_{\tilde{\pi}}(eB)/m_{\pi}(0)$ is almost identical to that
obtained when the couplings $g$ and $g_{v}$ are kept constant. If one allows
for different $B$ dependences for $g$ and $g_{v}$ it is possible to improve
on the agreement with LQCD results for this ratio. However, this implies a
rather strong enhancement in the masses of $S_{z}=\pm 1$ vector meson
states, leading to a rather large discrepancy with LQCD results in
Ref.~\cite{Bali:2017ian}.

For simplicity, in the present work we have not taken into account the axial
vector interactions. The influence of these degrees of freedom in the
magnetic field dependence of light neutral meson masses, and, in particular,
on the ratio $r_{\pi}$, is certainly an issue that deserves further
investigation. It would be also interesting to study the effect of the
inclusion of quark anomalous magnetic moments. We expect to report on these
issues in future publications.

\section*{Acknowledgements}

We are grateful to M.F.\ Izzo Villafa\~ne for helpful discussions at the
early stages of this paper. This work has been partially funded by CONICET
(Argentina) under Grant No.\ PIP17-700, by ANPCyT (Argentina) under Grants
No.\ PICT17-03-0571 and PICT19-0792, by the National University of La Plata
(Argentina), Project No.\ X284, by Ministerio de Ciencia e Innovaci\'on and
Agencia Estatal de Investigaci\'on (Spain) MCIN/AEI/10.13039/501100011033
and European Regional Development Fund Grant PID2019- 105439 GB-C21, by EU
Horizon 2020 Grant No.\ 824093 (STRONG-2020), and by Conselleria de
Innovaci\'on, Universidades, Ciencia y Sociedad Digital, Generalitat
Valenciana, GVA PROMETEO/2021/083.

NNS would like to thank the Department of Theoretical Physics of the
University of Valencia, where part of this work was carried out, for their
hospitality within the visiting professor program of the University of
Valencia.

\section*{Appendix A: Regularized $B=0$ polarization functions}

\newcounter{eraAA}
\renewcommand{\thesection}{\Alph{eraAA}}
\renewcommand{\theequation}{\Alph{eraAA}\arabic{equation}}
\setcounter{eraAA}{1} \setcounter{equation}{0} 
\label{appa}

In this appendix we give the expressions for the regularized
$B=0$ pieces of the polarization functions, $\hat
{\jnomatrix}^{0,{\rm reg}}_{M M'}(q)$, defined within the MFIR
scheme. As stated, it can be easily seen that these are zero for
$M\neq M'$, while for $M=M'$ one has
\begin{eqnarray}
\hat {\jnomatrix}_{\pi^a \pi^a}^{0,\rm reg}(q) & = & -  N_c \sum_f \Big[ I_{1f} + q^2 I_{2f}(q^2) \Big] \ ,
\nonumber \\
\hat {\jnomatrix}_{\rho_\mu^a \rho_\nu^a}^{0,\rm reg}(q) &=& \ \frac{2N_c}{3}\, \sum_f \Big[(2M_f^2-q^2)I_{2f}(q^2)-2M_f ^2I_{2f}(0)\Big]\,
\Big(\delta_{\mu\nu}-\frac{q_\mu q_\nu}{q^2}\Big)\ .
\label{b0reg}
\end{eqnarray}
Here, the integrals $I_{1f}$ and $I_{2f}(q^2)$ are defined as
\begin{eqnarray}
I_{1f} \ = \ 4 \int_p \ \frac{1}{M_f^2+p^2}\ ,
\qquad \qquad
I_{2f}(q^2) \ = \ -\,2\int_p \ \frac{1}{(M_f^2+p_+^2)\,(M_f^2+p_-^2)}\ ,
\label{eqI1eI2}
\end{eqnarray}
with $p^\pm = p \pm q/2$. Within the 3D-cutoff regularization scheme used in
this work, the first of these integrals is given by
\begin{equation}
I_{1f} \ = \ \dfrac{1}{2 \pi^2} \left[ \Lambda^2\, r_{\Lambda f} +
M_f^2 \ln\left( \dfrac{M_f}{\Lambda\, ( 1 + r_{\Lambda f})}\right) \right] \ ,
\label{I1freg}
\end{equation}
where we have defined $r_{\Lambda f} \equiv \sqrt{1+ M_f^2/\Lambda^2}$. In
the case of $I_{2f}(q^2)$, we note that in order to determine the meson
masses, the external momentum $q$ has to be extended to the region $q^2 <
0$. Hence, we find it convenient to write $q^2=-m^2$, where $m$ is a
positive real number. Then, within the 3D-cutoff regularization scheme, the
regularized real part of $I_{2f}(-m^2)$ can be written as
\begin{align}
{\rm Re}\left[ I_{2f}(-m^2)\right] = -\dfrac{1}{4\pi^2} &
 \left[ \arcsinh \left(\dfrac{\Lambda}{M_f}\right) - F_f\right]  \ ,
\end{align}
where
\begin{eqnarray}
F_f \ = \
\left\{
\begin{array}{ll}
  \sqrt{4 M_f^2/m^2 -1} \ \arctan \left( \dfrac{1}{r_{\Lambda f}\sqrt{4 M_f^2/m^2 -1}} \right) & \mbox{\ \ if \ \ } m^2 < 4 M_f^2 \\
  \sqrt{1-4 M_f^2/m^2}  \ \arccoth \left( \dfrac{1}{r_{\Lambda f}\sqrt{1-4 M_f^2/m^2}} \right) & \mbox{\ \ if \ \ }
  4 M_f^2 < m^2 < 4(M_f^2+\Lambda^2) \\
  \sqrt{1-4 M_f^2/m^2}  \ \arctanh \left( \dfrac{1}{r_{\Lambda f}\sqrt{1-4 M_f^2/m^2}} \right) & \mbox{\ \ if \ \ }
  m^2 > 4(M_f^2+\Lambda^2)
\end{array}
\right. \ . \nonumber
\end{eqnarray}
For the regularized imaginary part we get
\begin{equation}
{\rm Im} \left[ I_{2f}(-m^2)\right] \ = \
\left\{
\begin{array}{cl}
  -\dfrac{1}{8 \pi} \sqrt{1-4 M_f^2/m^2} & \mbox{\ \ if \ \ } 4 M_f^2 < m^2 < 4(M_f^2+\Lambda^2) \\
 \rule{0cm}{0.79cm} 0 & \mbox{\ \ otherwise}\\
\end{array}
\right. \ .
\label{ima2}
\end{equation}

\section*{Appendix B: Integrals $I_{nf}^{\rm mag}(-m^2)$ for $m > 2M_f$}

\newcounter{eraB}
\renewcommand{\thesection}{\Alph{eraB}}
\renewcommand{\theequation}{\Alph{eraB}\arabic{equation}}
\setcounter{eraB}{2} \setcounter{equation}{0} 

The expressions for the integrals $I_{nf}^{\rm mag}(-m^2)$ for
$n=2,\dots ,5$ given in Eqs.~(\ref{inf}) are only valid when $m <
2M_f$. For $m > 2 M_f$, it happens that the corresponding
integrands can become divergent at some points within the
integration domain, leading to divergent integrals. However, we
can get finite results by considering the analytical extension of
the functions in Eqs.~(\ref{inf}). For this purpose it is worth
taking into account that the Feynman quark propagators originally
contain ``$i\epsilon$'' terms, which can be easily recovered in the
integrands of Eqs.~(\ref{inf}) through the replacement $M_f^2
\rightarrow M_f^2 - i\epsilon$ (note that this implies the
replacement $\bar x_f \rightarrow \bar x_f - i\epsilon$). Once
this is done, one can proceed by using the digamma recurrence
relation

\begin{equation}
\psi(x) = \psi(x+n+1) - \sum_{j=0}^n \dfrac{1}{x+j} \ ,
\end{equation}

and taking $\epsilon\rightarrow 0^+$ through a generalized version
of the Sokhotski-Plemelj formula [see e.g.\ Eq.~(A8) of
Ref.~\cite{Avancini:2021pmi}]. In this way, we find that for $m >
2 M_f$ the integrals $I_{nf}^{\rm mag}(-m^2)$, $n=2,\dots ,5$,
can be extended to
\begin{eqnarray}
I^{\,{\rm mag}}_{2f}(-m^2) &=& \frac{1}{8 \pi^2} \Bigg[ \int_0^1 dv \ \psi(\bar x_f + N + 1)
\nonumber \\
& &\qquad  - \ln x_f + 2 - 2 \beta_0 \arctanh \beta_0 + \frac{4
B_f}{m^2} \sum_{n=0}^N \frac{\alpha_n}{\beta_n} \arctanh\beta_n
\Bigg]
\nonumber \\
& & + \frac{i}{8 \pi} \Bigg[ \beta_0 - \frac{2 B_f}{m^2} \sum_{n=0}^N \frac{\alpha_n}{\beta_n} \Bigg]
\ ,
\\[4.mm]
I^{\,{\rm mag}}_{3f}(-m^2) &=& - \frac{Q_f M_f}{\pi^2 m} \left[
\frac{\arctanh \beta_0}{\beta_0} - i \frac{\pi}{2 \beta_0} \right]
\ ,
\\[4.mm]
I^{\,{\rm mag}}_{4f}(-m^2) &=& - I^{\,{\rm mag}}_{1f}\,  +\,
T_f^+(-m^2)\, +\,
T_f^-(-m^2) \nonumber \\
& & - \frac{m^2}{16 \pi^2} \Bigg[ 4\beta_0 (1-\frac{1}{3}\, \beta_0^2)
\arctanh \beta_0 + \left(\frac{7}{3}-\beta_0^2\right) \ln x_f + \frac{4}{3}\, \beta_0^2 - \frac{38}{9}
\Bigg] \nonumber \\
& &  + \,
 \frac{i\,m^2}{8\pi}\Bigg[\beta_0 (1-\frac{1}{3}\, \beta_0^2) - \theta(m-m_f^-)\,
 \frac{4B_f}{m^2}
 \sum_{n=0}^{N^-}\frac{(2\lambda - 2n - 1)}{r_n} \Bigg] \ ,
\label{i4f}
\end{eqnarray}
\begin{eqnarray}
I^{\,{\rm mag}}_{5f}(-m^2) &=& \frac{1}{8 \pi^2} \Bigg[ \int_0^1
dv \ (1-v^2) \ \psi(\bar x_f + N + 1)
\nonumber \\
& &\qquad - \frac{2}{3} \left( \ln x_f - \frac{8}{3} + \beta_0^2 + \beta_0 (3-\beta_0^2) \arctanh \beta_0 \right)
\nonumber
\\
& &\qquad  + \frac{4 B_f}{m^2} \sum_{n=0}^N
\frac{\alpha_n}{\beta_n} \left[ \beta_n +(1- \beta_n^2) \arctanh \beta_n  \right] \Bigg]
\nonumber \\
& & + \frac{i}{8 \pi} \Bigg[ \beta_0 (1-\frac{1}{3}\, \beta_0^2) - \frac{2
B_f}{m^2} \sum_{n=0}^N \frac{\alpha_n}{\beta_n} (1-\beta_n^2)
\Bigg] \ .
\end{eqnarray}
Here, we have used the definition $\alpha_n = 2 - \delta_{0n}$, together with
\begin{equation}
\beta_n = \sqrt{ 1 - \frac{4 (M_f^2 + 2n B_f)}{m^2} }\ , \qquad
r_n = \sqrt{1 - 4\lambda(2n+1) + 4\lambda^2\beta_0^2} \ , \qquad
N = \mbox{Floor}\left[ \lambda \beta_0^2 /2 \right] \ ,
\nonumber
\end{equation}
where $\lambda= m^2/(4 B_f)$. In the expression of $I^{\rm
mag}_{4f}(-m^2)$ the integral $I_{1f}^{\rm mag}$ is that given
by Eq.~(\ref{i1}), and we have introduced the functions
$T_f^\pm(-m^2)$ given by
\begin{eqnarray}
\!\!\!\!\!\!\!\!\! T^\pm_f(-m^2) &=&
  \frac{m^2}{32\pi^2}\int_0^1 dv \   ( v^2 \pm v/\lambda + \gamma) \
  \psi(\bar x_f + (1\pm v)/2 + \,\theta(m-m^\pm_f)\,(1 +N^\pm))
\nonumber \\
&&- \frac{B_f}{4\pi^2}\,\theta(m-m^\pm_f) \sum_{n=0}^{N^\pm} \left\{ 1 -
\frac{(2\lambda - 2n-1)}{r_n} \ln \left[
\frac{(2\lambda - r_n \pm 1)|r_n \pm 1|}{(2\lambda + r_n \pm 1)(r_n\mp 1)}
\right]\right\}\ ,
\label{Tt}
\end{eqnarray}
with $\gamma =2-\beta_0^2 = 1+4M_f^2/m^2$, and
\begin{equation}
m_f^- = M_f + \sqrt{M_f^2+2B_f}\ ,\qquad\qquad m_f^+ = 2
\sqrt{M_f^2+B_f} \ .
\label{emes}
\end{equation}
The integers $N^\pm$ have been defined as
\begin{equation}
N^- = \mbox{Floor} \left[ r_0^2/(8\lambda) \right]\ , \qquad\qquad
N^+ = \mbox{Floor} \left[ (\lambda\beta_0^2 -1)/2 \right] \ .
\end{equation}

\section*{Appendix C: A simplified model for the lowest state of the $S_z=0$ sector}

\newcounter{eraC}
\renewcommand{\thesection}{\Alph{eraC}}
\renewcommand{\theequation}{\Alph{eraC}\arabic{equation}}
\setcounter{eraC}{3} \setcounter{equation}{0} 

In this appendix we present a simplified model to analyze the mass and
composition of the lowest state of the $S_z=0$ meson sector. As seen in
Sec.~\ref{sec_sz0} (see the discussion concerning Fig.~\ref{Fig4}), the mass
of this state, while almost independent of the value of $\alpha$, is
significantly affected by the existence of a mixing between pseudoscalar
and vector meson states. Thus, to simplify the analysis we consider the
case $\alpha=0.5$, in which the relevant basis is only composed by the
states $\pit$, $\rhotpar$ and $\rhocpar$. In addition, one has
$M_u=M_d\equiv M$ for any value of $eB$. Assuming as in the main text
$g_{v_0}=g_{v_3}=g_{v}$, it is easy to see that the ratio between the
off-diagonal $\pit\rhotpar$ and $\pit\rhocpar$ mixing matrix elements is
given by $G_{\parallel\pit\rhot}/G_{\parallel\pit\rhoc} =
(B_u-B_d)/(B_u+B_d) = 1/3$. Hence, to simplify the problem even further, in
what follows we only consider the $\pit$ - $\rhocpar$ system (see, however,
discussion at the end of this appendix). To check whether we are capturing
the main effect of pseudoscalar-vector meson mixing on the $\tilde \pi$
mass it is useful to consider the ratio $r_\pi =
m_{\tilde\pi}(eB)/m_\pi(0)$. Assuming that $g$ and $g_v$ do not depend on
the magnetic field, and taking $\alpha = 0.1$, for $eB = 1$~GeV$^2$ we get
$r_\pi=0.32$ for the full $\pic$ - $\pit$ - $\rhocpar$ - $\rhotpar$ system,
to be compared with the values $r_\pi=0.38$, obtained when we consider only
the $\pit$ - $\rhocpar$ system, and $r_\pi=0.92$, obtained for the case in
which there is no mixing at all. These values clearly support our
approximation of the full system by the much simpler $\pit\rhocpar$ one. It
should be stressed that even in this simplified situation the lowest
mass state is still found to be strongly dominated by the $\pit$
contribution. In fact, for $eB = 1$~GeV$^2$ we get $c^{(1)}_{\rhocpar} =
-0.083$, close to the value $-0.0797$ obtained for the full system (see
Table~\ref{taba}). Defining a mixing angle $\theta$ by $\tan \theta =
c^{(1)}_{\rhocpar}/c^{(1)}_{\pit}$, this implies $\theta\simeq - 5^0$.

{  The strong dominance of the $\pit$ contribution to the
$\tilde\pi$ state suggests that one should be able to determine
the mixing effect on $m_{\tilde\pi}$ using first order
perturbation theory.} On the other hand, this appears to be in
contradiction with the aforementioned significant reduction of the
$\tilde\pi$ mass. To get a better understanding of the situation,
it is convenient to carry out some further approximations. The
relevant mixing matrix elements to be considered are
\begin{eqnarray}
\!\!\!\!\! G_{\parallel\pit\pit} &=& \frac{1}{2g} - N_c \sum_f
\left[ (I_{1f} + I_{1f}^{\rm mag}) - m^2 (I_{2f}(-m^2) + I_{2f}^{\rm mag}(-m^2))\right]\ ,
\nonumber \\
\!\!\!\!\! G_{\parallel\pit\rhoc} &=& \frac{i\,N_c\, s_B B_e\,
\arctan\Big(1/\sqrt{4 M^2/m^2 - 1}\Big)}{2\pi^2\,\sqrt{1-m^2/4 M^2}} \ ,
\nonumber \\
\!\!\!\!\! G_{\parallel\rhoc\rhoc} &=& \frac{1}{2g_{v}} + N_c
\sum_f  \left[ \frac{2}{3}\,
\Big[(2M^2+m^2)I_{2f}(-m^2)-2M^2I_{2f}(0)\Big]\, + m^2
I_{5f}^{\rm mag}(-m^2) \right] \ ,
\end{eqnarray}
where we have denoted $B_e = |eB|$ and $s_B=\mbox{sign}(B)$. For
the $\tilde\pi$ state, we have $m^2/(4 M^2)\ll 1$ (for our
parametrization we find $m^2/(4 M^2)\approx 0.03$ at vanishing
magnetic field, and even a smaller value at $eB= 1$\ GeV$^2$).
Thus, we can obtain a good approximation to these matrix elements
by expanding up to ${\cal O}(m^2/4M^2)$. In this way we get a
mixing matrix of the form
\begin{eqnarray}
G_\parallel^{(\pi_3\rho_0)} \ = \ \left(%
\begin{array}{cc}
\alphagen - \betagen\, m^2 &  i\,c\,m \\
 -i\,c\,m & \gammagen - \betagen'\, m^2 \\
\end{array}%
\right) \ ,
\label{matrix}
\end{eqnarray}
where
\begin{equation}
\alphagen = \frac{m_c}{2 g M} \ ,\quad\quad c = \frac{N_c \, s_B B_e}{4 \pi^2 M} \ ,
\quad\quad \gammagen = \frac{1}{2 g_{v}}\ , \quad\quad
\betagen' =\frac{2\betagen}{3}  + \frac{N_c}{18\pi^2}
\frac{\Lambda^3}{(\Lambda^2 + M^2)^{3/2}}
\label{eqcoef}
\end{equation}
and
\begin{equation}
\betagen  = \frac{N_c}{8 \pi^2} \left[ 4\, {\rm arcsinh}
\frac{\Lambda}{M} - \frac{4 \Lambda}{\sqrt{\Lambda^2 + M^2}} -
\frac{  B_e}{M^2} + \ln \frac{9 M^4}{8 B_e^2} - \psi\left(\frac{3
{  M^2}}{4 B_e}\right) -
  \psi\left(\frac{3 M^2}{2 B_e}\right) \right]\ .
\label{eqcoefa}
\end{equation}
We note that here the gap equation has been used to get the expression for
$\alphagen$. Given the model parameters, these coefficients can be easily
computed for a given value of $B_e$.

Keeping terms up to the leading order in $m^2$, one gets in this way
\begin{equation}
m^2_{\tilde\pi} \ = \ \frac{a\, d}{a\, b' + b\, d + c^2} \ .
\end{equation}
In addition, it can be seen that $\alphagen/\gammagen = (g_{v_0}/g)
(m_c/M)\ll 1$, and consequently $ab'\ll bd$. Using this approximation we
obtain
\begin{equation}
m_{\tilde\pi} \ = \ \frac{\bar m_{\tilde\pi}}{\sqrt{1 + c^2/(\betagen\, \gammagen)}} \ ,
\label{eqapp}
\end{equation}
where $\bar m_{\tilde\pi} = \sqrt{\alphagen/\betagen}$ is the mass of the
lightest state if there is no mixing at all. Within the same approximation,
the coefficient of the $\rhocpar$ piece of the lightest state is given by
\begin{equation}
c^{(1)}_{\rhocpar} \ = \ -\frac{m_{\tilde\pi}\, c}{\gammagen}\ .
\label{eqcoe}
\end{equation}

The numerical values for the above quantities can be calculated from
Eqs.~(\ref{eqcoef}), (\ref{eqcoefa}). For a large magnetic field
$eB=1$~GeV$^2$, assuming that the coupling constants are independent of $B$,
we get $r_\pi=0.39$ and $c^{(1)}_{\rhocpar}=-0.084$, in excellent agreement with
the results quoted above for the $\pit\rhocpar$ system. This confirms the
validity of the approximations made so far.

It is also interesting to note that the expression for $m_{\tilde\pi}$ given in
Eq.~(\ref{eqapp}) implies
\begin{eqnarray}
b\,(m_{\tilde\pi}^2 - \bar m_{\tilde\pi}^2) \ = \ -\,\frac{m_{\tilde\pi}^2 \,
c^2}{\gammagen}\ .
\end{eqnarray}
This expression, together with the one for $c^{(1)}_{\rhocpar}$ in
Eq.~(\ref{eqcoe}), are the relations that one would obtain from a
first-order perturbation analysis of the system described by the matrix in
Eq.~(\ref{matrix}) when $\gammagen \gg \alphagen + (\betagen' - \betagen)
m^2$, a condition that is always well satisfied in our case. One can
observe that the somewhat unexpectedly ``large'' value of the mass shift
arises from the small value of the coefficient $\betagen$, which is found to
be about 0.034 for $eB= 1$~GeV$^2$ (assuming $B$-independent couplings). In
a conventional eigenvalue problem, one would have $\betagen = 1$.

Finally, we note that the effect on this game of the $\rhotpar$ meson, so
far neglected, can be easily taken into account at this stage. Since, as
shown above, the $G_{\parallel\pit\rhoc}$ matrix element can be treated
perturbatively, and $G_{\parallel\pit\rhot}$ is even 3 times smaller, one
can account for the $\rhotpar$ meson just replacing the factor $c^2$ in
Eq.~(\ref{eqapp}) by $10/9\, c^2$. The resulting expression for
$m_{\tilde\pi}$ can be rewritten as
\begin{equation}
m_{\tilde\pi} \ = \ \frac{\bar m_{\tilde\pi}}{\sqrt{1 +
\kappa\, \bar m_{\tilde\pi}^2\, B_e^2/M}}\ \; ,
\label{eqappb}
\end{equation}
where $\kappa = 5 N_c^2\, g\, g_{v}/(18 \pi^4 m_c)$. To obtain this
expression we have made use of Eq.~(\ref{eqcoef}) together with the relation
$\betagen = \alphagen/\bar m_{\tilde\pi}^2 = m_c/(2 g M \bar m_{\tilde\pi}^2)$.

It should be emphasized that although the numerical values quoted in this
appendix correspond to the case in which the couplings $g$ and $g_v$ are
kept fixed, Eq.~(\ref{eqappb}) can be shown to be approximatively valid also
when they depend on $B$ as considered in Sec.~\ref{bdependent}.

\end{document}